\documentclass{aa}  

\usepackage{graphicx}
\usepackage{txfonts}
\begin{document}

   \title{A re-classification of  Cepheids in the {\bf \it Gaia} Data Release 2}

   \subtitle{Period-Luminosity and Period-Wesenheit relations in the
     {\bf \it Gaia} passbands}

   \author{V. Ripepi
          \inst{1}
          \and
          R. Molinaro 
          \inst{1}
          \and 
          I. Musella  
          \inst{1}
          \and 
          M. Marconi  
          \inst{1}
          \and 
          S. Leccia 
          \inst{1}
          \and 
          L. Eyer 
          \inst{2}
          }

   \institute{INAF-Osservatorio Astronomico di Capodimonte, Via
     Moiariello 16, 80131, Naples, Italy\\
              \email{vincenzo.ripepi@inaf.it}
         \and 
             Department of Astronomy, University of Geneva, Ch. des Maillettes 51, CH-1290 Versoix, Switzerland\\
             %\email{gisella.clementini@inaf.it}
             }

   \date{...}

% \abstract{}{}{}{}{} 
% 5 {} token are mandatory
 
  \abstract
   {Classical Cepheids are the most important primary indicators for
     the extragalactic distance scale. Establishing the precise zero
     points of their Period-Luminosity and Period-Wesenheit ($PL/PW$) relations has profound
     consequences on the estimate of $\rm H_0$. Type II Cepheids are
     also important distance indicator and tracers of old stellar populations.}
   {The recent Data Release 2 (DR2) of the {\it Gaia} Spacecraft includes
     photometry and parallaxes for thousands of classical and Type II 
     cepheids. We aim at reviewing the classification of {\it Gaia} DR2
     Cepheids and to derive precise $PL/PW$ for
     Magellanic Cloud (MCs) and Galactic Cepheids.}
   {Information from the literature and the {\it Gaia} astrometry and
     photometry are adopted to assign DR2 Galactic
     Cepheids to the classes: Classical, Anomalous and Type II Cepheids.}
   { We re-classified the DR2 Galactic Cepheids and derived new precise
     $PL/PW$ relations in the {\it Gaia} passbands for the MCs and Milky
     Way Cepheids.
We investigated for the first time the dependence on metallicity of
the $PW$ relation for Classical Cepheids in the {\it Gaia} bands, finding non-conclusive
results.}  
   {According to our analysis, the zero point of the {\it Gaia} DR2 parallaxes as estimated from
     Classical and Type II Cepheids seems to be likely
     underestimated by $\sim$0.07 mas, in full agreement with recent literature. The
     next {\it Gaia} data releases are expected to fix this zero point
     offset to eventually allow a determination of $\rm H_0$ to
     less than 1\%.}


   \maketitle
%
%-------------------------------------------------------------------

\section{Introduction}

Classical Cepheids (CCs) are the most important primary distance
indicators for the cosmic distance scale \citep[see e.g.][]{Riess2016,Riess2018a}, 
due to their characteristic Period–Luminosity ($PL$) and
Period-Wesenheit ($PW$) relations \citep{Leavitt1912,Madore1982,Caputo2000}.

In conjunction with secondary distance indicators such as SNIa, the
CCs provide an estimate of  $\rm H_0 \sim 73.48\pm1.66$ km/sec/Mpc with 2.3\% of claimed uncertainty  
\citep[][]{Riess2018b}. 

However, there is a tension at 3.4-3.7 $\sigma$ with
$\rm H_0 \sim 66.93\pm0.62$ km/sec/Mpc obtained from the analysis of the
cosmic microwave background plus $\Lambda$CDM
\citep[][]{Planck2016,Riess2018a,Riess2018b}.

To reconcile the inconsistency between these values, we need more
accurate calibrations of the different steps of the cosmic distance
ladder. In first place we have to check the calibration of 
slopes/zero points of the $PL/PW$ relations used
for CCs that at moment rely on a handful of objects with accurate 
Hubble Space Telescope (HST) parallaxes \citep[][]{Riess2018a}.

In this context a great help can be given by the measures of the 
astrometric spacecraft \textit{Gaia} \citep{GaiaPrusti2016} that is collecting
repeated multi-band photometric and astrometric data  of sources all over the
sky to a limiting magnitude of about $G \sim 20.7$ mag. 

The \textit{Gaia} Data Release 2 (DR2) \citep[see][for a detailed description
of the content of the release]{GaiaBrown2018} has published
photometry in the three \textit{Gaia} pass-bands $G$, $G_{BP}$ and
$G_{RP}$, as well as astrometry and radial velocity data obtained
during the initial 22 months of data collection.  

The multi-epoch \textit{Gaia} data permitted the study of an
unprecedented number of variable stars of different types
\citep[for details see][]{Holl2018}. In particular, \citet{Clementini2018} discussed 
the pipeline of the {\it Cepheid\&RRLyrae Specific Object Studies (SOS)}
used to measure period(s), intensity-averaged $G$, $G_{BP}$ and
$G_{RP}$ magnitudes and amplitudes of pulsation for a sample of 140,784 RR Lyrae and 9,575
Cepheids. Among the latter,  3,767, 3,692, and 2,116  are Cepheids belonging to the
Large Magellanic Cloud (LMC), Small Magellanic Cloud (SMC) and All
Sky  sample, respectively. The latter sample, consist essentially in candidate
Cepheids belonging to our Galaxy. In the following we will refer to these stars as the
Milky Way (MW) sample.  As a result of a complex 
concomitant factors (automatic procedure, inaccurate parallaxes etc.)
the MW sample is thought to be significantly contaminated by
non-Cepheid types of variable stars \citep[see][for details]{Clementini2018}.
Therefore, the main scope of this paper is to provide a 
detailed re-classification of the objects classified as Cepheids (of
different types, see below) in
\citet{Clementini2018}, providing a comparison with the classification
in the literature. We also aim at calculating empirical $PL/PW$ relations in the Gaia
pass-bands for the LMC/SMC and MW for future uses. 

%-------------------------------------- Two column figure (place early!) 
   \begin{figure*}
   \centering 
   \includegraphics[width=12cm]{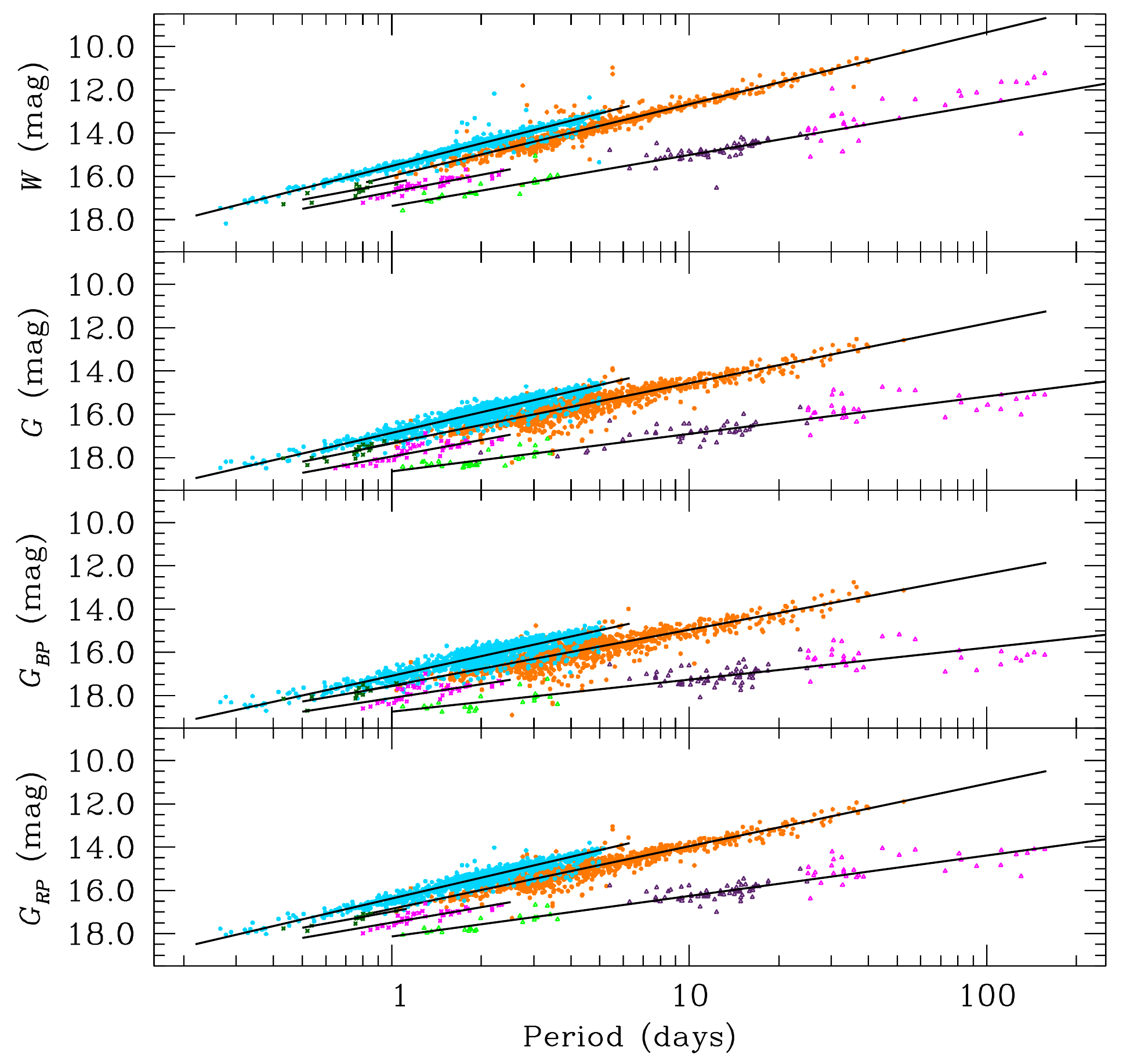}
   \caption{PL/PW relations for the LMC in the form $mag=\alpha +\beta
     \log P$. From top to bottom panel $mag$ is represented by the
     apparent $W$, $G$, $G_{BP}$, and $G_{RP}$ magnitudes, respectively.
Orange filled circles: DCEP\_F; cyan filled circles: DECP\_1O; magenta four-starred 
symbols: ACEP\_F; dark green four-starred symbols: ACEP\_1O; green open triangles: BLHER; violet open triangles: WVIR; magenta open 
triangles: RVTAU.}
              \label{plLMC}%
    \end{figure*}

   \begin{figure*}
   \centering 
   \includegraphics[width=12cm]{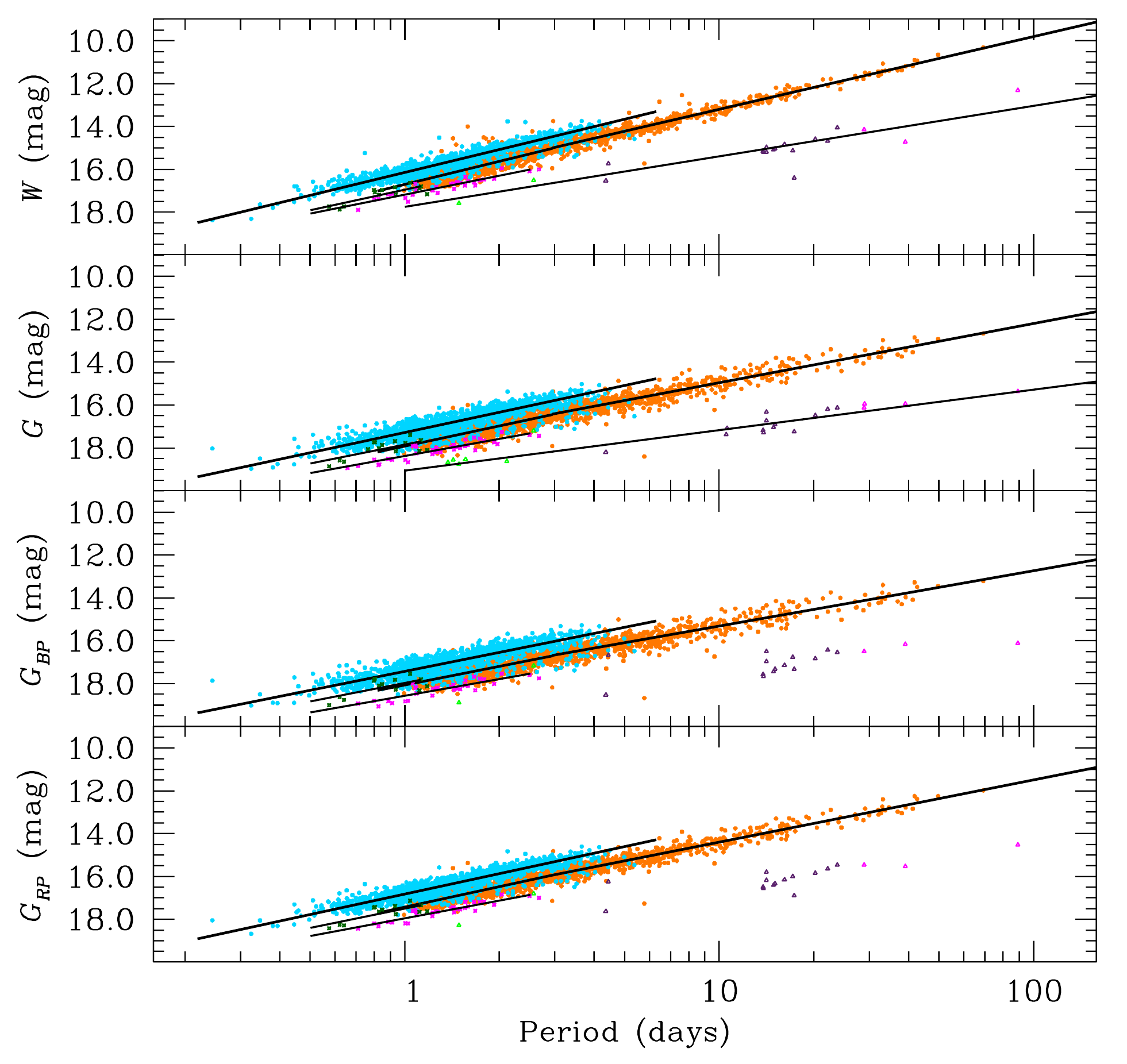}
   \caption{As in Fig.~\ref{plLMC} but for the SMC. 
              \label{plSMC}}%
    \end{figure*}
%
%

%--------------------------------------------------------------------

Before proceeding, we recall that we distinguish three types of
Cepheid variables: Classical Cepheids (DCEPs), Anomalous Cepheids
(ACEPs) and Type II Cepheids (T2CEPs). The latter type is usually
sub-divided into three subclasses, BL Her (BLHER), W Vir (WVIR) and RV
Tau (RVTAU), in order of increasing periods. DCEPs and ACEPs are known
to pulsate in different modes. In this paper we consider DCEPs
pulsating in the fundamental, first overtone, second
overtone\footnote{Note that second overtone pulsators were not
  classified by the {\it Cepheid\&RRLyrae Specific Object
    Studies (SOS)} pipeline} and
multiple mode: we name these variables as DCEP\_F, DCEP\_1O, DCEP\_2O,
DCEP\_MULTI, respectively. Similarly, for ACEPs we distinguish objects
pulsating in the fundamental (ACEP\_F) and first overtone modes
(ACEP\_1O). For a detailed description of these classes of variability
and their evolutionary status we refer the reader to e.g. the recent
textbook by \citet{Catelan2015}.

%Secondly, we aim at providing accurate PL/PW
%relations for a selected sample of MW CCs for which good 
%Gaia DR2 parallaxes, $VJHK_S$ photometry, reddening, metallicity and 
%binary information are available. These relations, especially those
%including Near-Infrared (NIR) bands are important for
%the extragalactic distance scale also at the light of the fact that
%next generation instruments such as Jams Webb Space Telescope (JWST)
%and the Extremely Large Telescope (ELT) will observe in these
%photometric regime.  

The manuscript is organised as follows:
in Section 2 we derive empirical $PL/PW$ relations for all type of
Cepheids in the LMC/SMC; in Section 3 we present the result of the
literature search; in Section 4 we carry out the re-classification of
the MW Cepheids and calculate their $PL/PW$ relations; a brief
summary closes the paper.

%-------------------------------------- Two column figure (place early!) 
   \begin{figure*}
   \centering 
\vbox{
 \includegraphics[width=9cm]{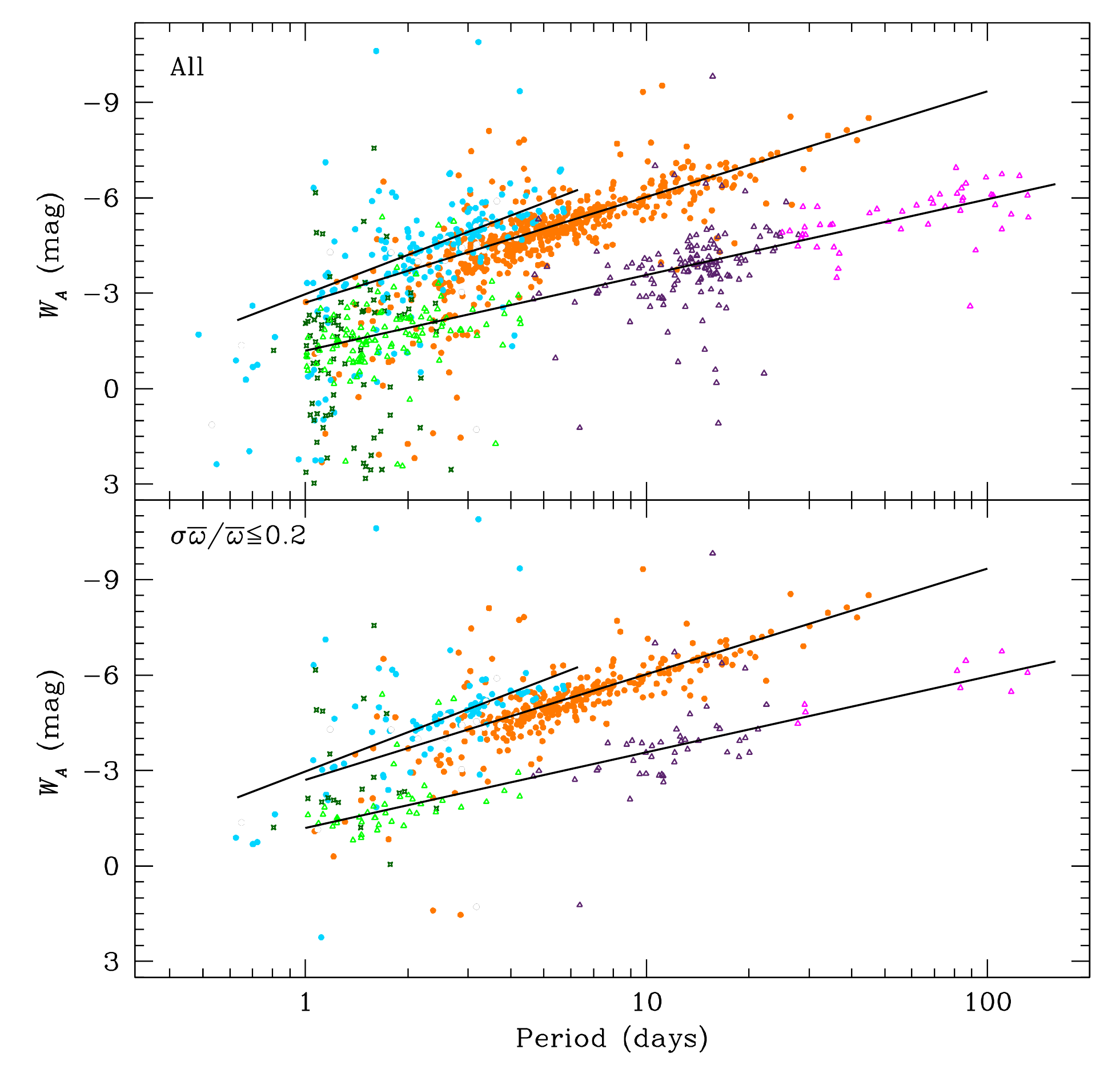}
 \includegraphics[width=9cm]{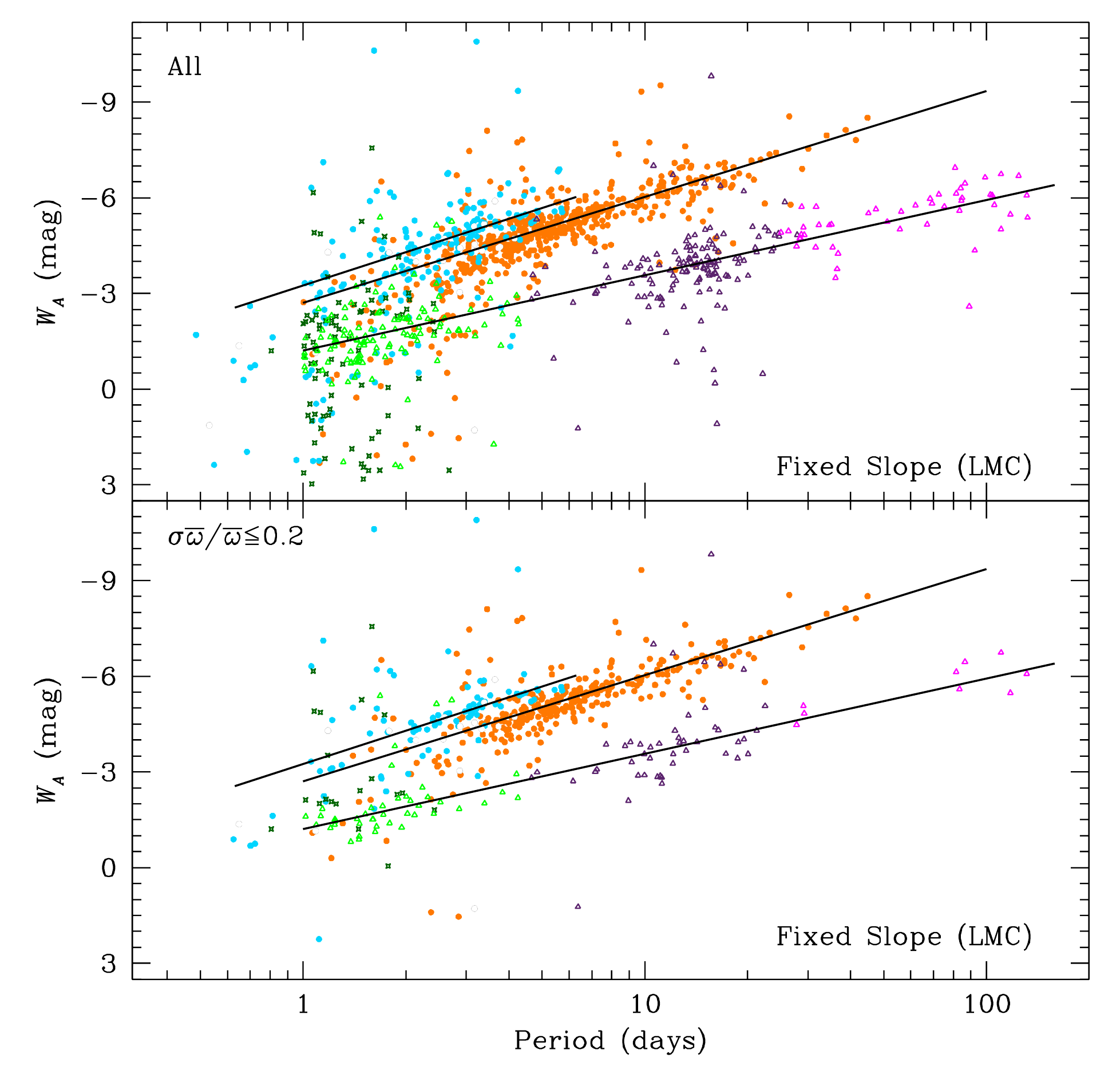}
}
   \caption{$PW$ relations for the re-classified MW sample. 
Orange filled circles: DCEP\_F; cyan filled circles: DECP\_1O; magenta four-starred 
symbols: ACEP; green open triangles: BLHER; violet open triangles: WVIR; magenta open 
triangles: RVTAU. Top and low panels show the 
complete sample and that with relative error on parallax better than 
20\%, respectively. Solid lines represent the least-square fit to the 
data obtained with the ABL method (see text). The $PW$ relations are of 
the form $W_A$=$\alpha +\beta \log P$. Left panels shows the $PW$ relations 
obtained with $\beta$ coefficient treated as unknown parameter in 
Eq.~\ref{eqABL}. In the right panels the $\beta$ coefficient is fixed
and equal to that obtained from the LMC.}
              \label{figPLMW}%
    \end{figure*}

%-------------------------------------- Two column figure (place early!) 
   \begin{figure*}
   \centering 
\hbox{
   \includegraphics[width=9cm]{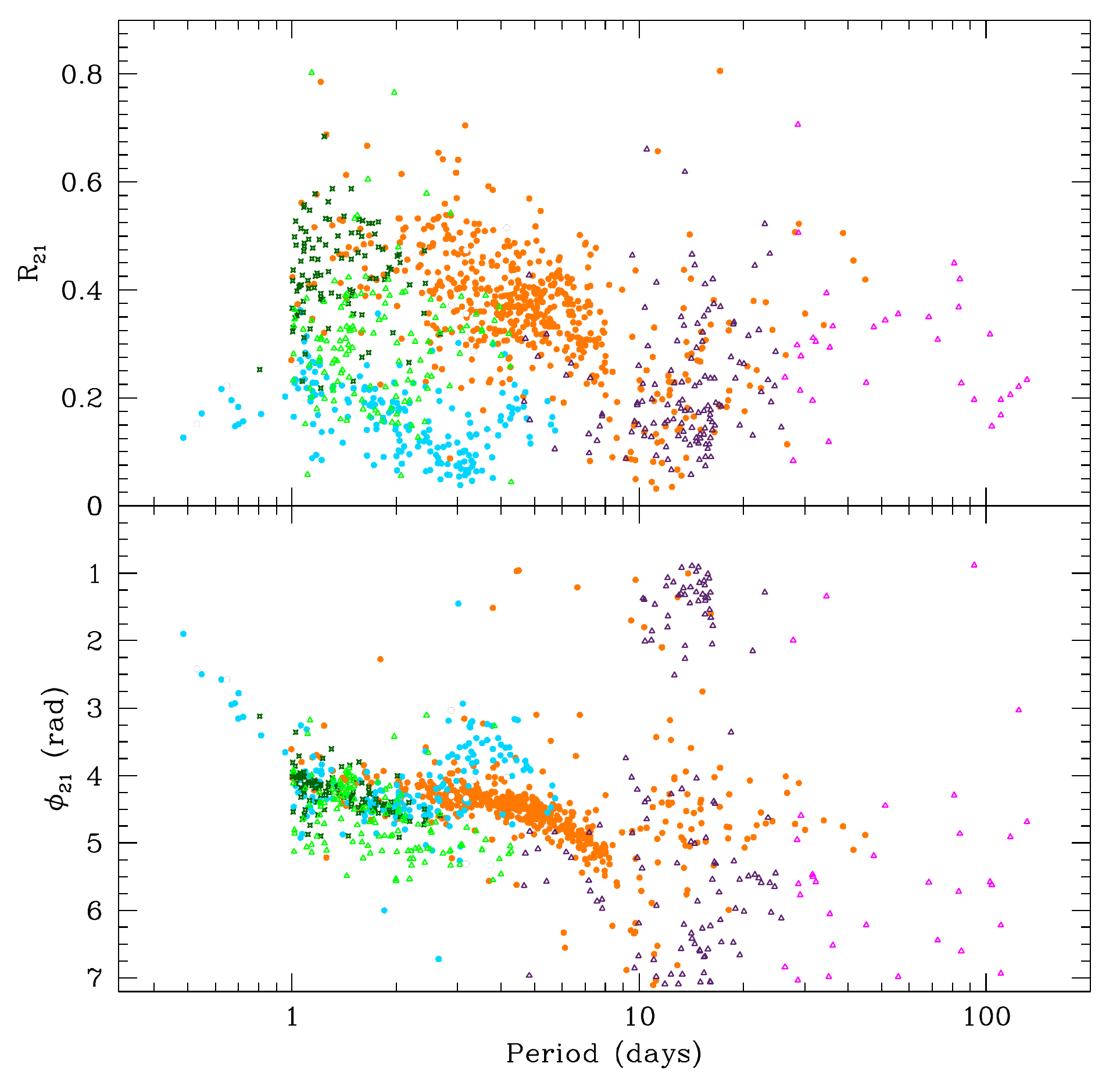}
   \includegraphics[width=9cm]{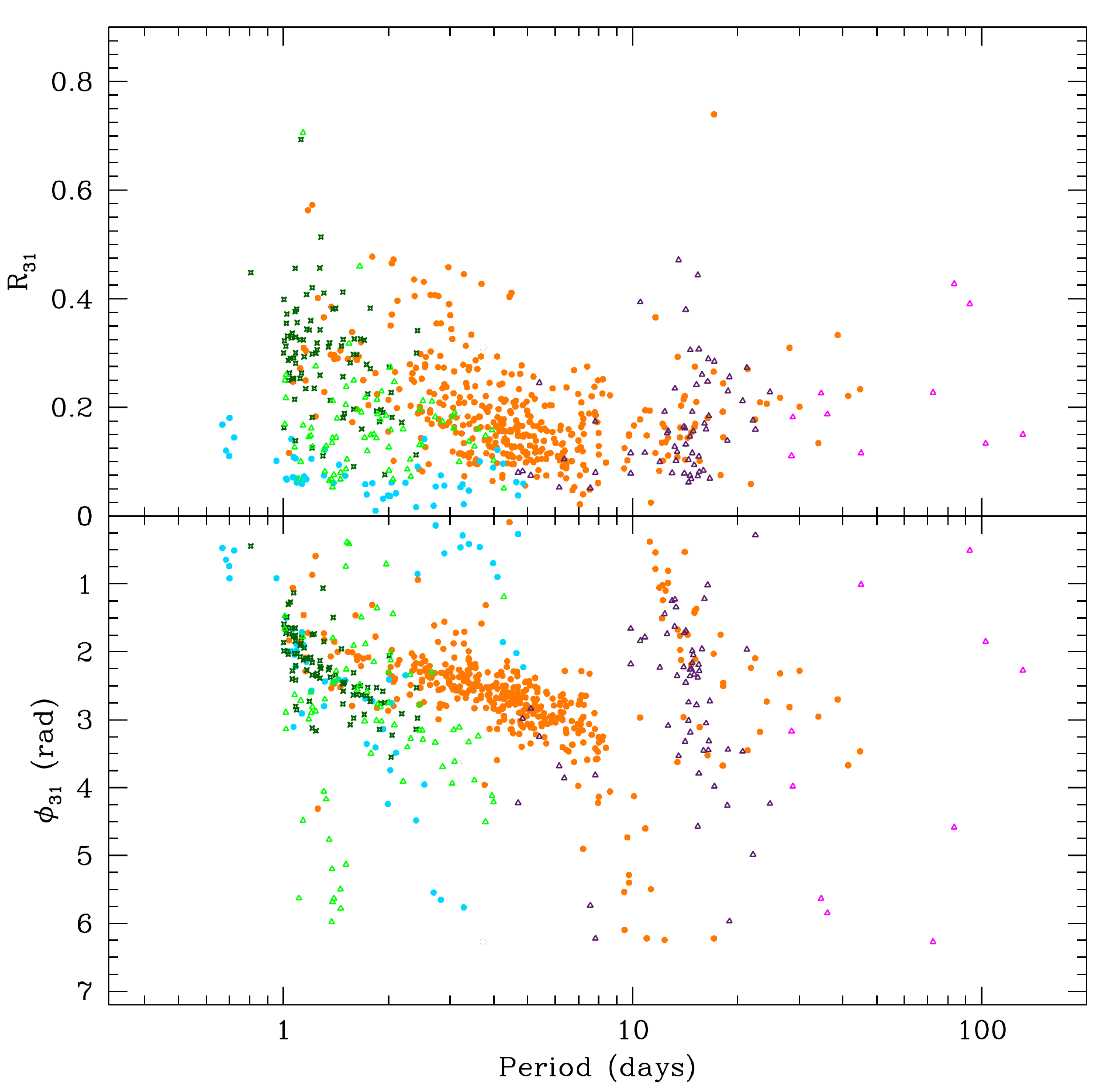}
}
   \caption{Fourier parameters for the re-classified 
     objects. Color-code as in Fig.~\ref{figPLMW}}
              \label{figFourier}%
    \end{figure*}

\section{\textit{\bf Gaia} DR2 Cepheids in the Magellanic Clouds}
\label{sectMCs}

Before facing the task of re-classifying the MW Cepheids in DR2, it is
first useful to analyse the DR2 output for the MCs Cepheids. 

In \citet{Clementini2016,Clementini2018}, i.e. for DR1 and
DR2, respectively, we used $PL/PW$ relations derived from
OGLE-III\footnote{http://ogle.astrouw.edu.pl/}
(Optical Gravitational Lensing Experiment) $V,I$
photometry transformed in $G,G_{RP}$ bands on the basis of
\citet{Jordi2010} predicted color transformations. 
It is then important to derive accurate $PL/PW$
relations for the different types of Cepheids in the {\it Gaia} passbands
directly from the actual data. 

In DR2, 3,767 and 3,692 Cepheids of all
types in the LMC and SMC were released, respectively \citep[see Table
2 of][for full details]{Clementini2018}.  These samples were
  complemented with 61 and 73 Cepheids coming from the MW sample, but
  actually belonging to the LMC and SMC, respectively, as shown in Sect.~\ref{altriSistemi}
  (see also Tab.~\ref{inAltriSistemi}). 
For DCEPs we first discarded multiple pulsators and used only objects
with reliable values of the three $G,G_{BP},G_{RP}$ bands. We were then
left with 1,624 and 1,207 DCEP\_F and DCEP\_1O pulsators  in the LMC, as well
as 1,772 and 1,368 DCEP\_F and DCEP\_1O pulsators in the SMC, respectively.
We did no attempt to correct the classification of these objects
because it had been already demonstrated that it is very accurate
\citep[see Fig. 41 in][]{Clementini2018}.

Secondly, we decided to use a different formulation of the Wesenheit
magnitude with respect to that used in {\it Gaia} DR1 and DR2, 
involving only $G$ and $G_{RP}$ bands
\citep[see][for details]{Clementini2016,Clementini2018}. The new
formulation is the following: 

\begin{equation}
W(G,G_{BP},G_{RP})=G-\lambda(G_{BP}-G_{RP})
\label{eqWesenheit}
\end{equation} 

\noindent
where $\lambda=A(G)/E(G_{BP}-G_{RP})$. 
Empirically, it is known that the value of $\lambda$ is of the order of 2
 over a wide range of effective temperatures,
including those typically spanned by Cepheids \citep{Andrae2018}. To obtain a more precise value, we adopted the
synthetic photometry by \citet{Jordi2010}, that provides the value of
$\lambda$ as a function of effective temperature, gravity and
metallicity. We selected the ranges of these parameters typical for
Cepheids  (i.e. 4500 $<T_{eff}<$7000 K; 0.5$< \log g<$3.5 dex;
-1$<[Fe/H]<$+0.5 dex) and then averaged out the selected
values, obtaining $\lambda=1.95\pm0.05$, very close to the
\citet{Andrae2018} result. 

We tested this Wesenheit magnitude on DCEPs
in the LMC, which are known to show  very tight $PW$ relations in all bands \citep[see
e.g.][in the optical and near infrared,
respectively]{Sos2017a,Ripepi2012}. After a few experiments, we
realised that the least-square fit to the data gave a tighter $PW$ relation
(smaller scatter) if the $\lambda$ value was slightly decreased to
1.90 (with an uncertainty estimated of the order of 0.05, by looking
at the value of $\lambda$ that produced an increase in the
dispersion). 
Hence, in the following we decided to use the following 
Wesenheit magnitude:

\begin{equation}
W(G,G_{BP},G_{RP})=W=G-1.90(G_{BP}-G_{RP}),
\label{wdefinition}
\end{equation}

\noindent
where $G$, $G_{BP}$, $G_{RP}$ are the magnitudes in the {\it Gaia}
bands. 
In comparison to that used in \citet{Clementini2016,Clementini2018},
the new formulation has
the advantage to be linear in the color term and to provide smaller
dispersions in the $PW$ relations.

Apart from the $PW$ relation, we also derived individual $PL$s 
for the $G,G_{BP},G_{RP}$ bands. We did no attempt to correct for
extinction because no reliable individual reddening estimate is present
in the literature \citep[see e.g.][for a discussion on the
uncertainties in the individual reddening value for
DCEPs]{Gieren2018}. In any case, the average foreground reddening values in LMC
and SMC are known to be small, of the order of
$E(B-V)\approx0.08$ and $0.04$ mag, respectively  (see e.g. the values from NASA/IPAC
Extragalactic Database - NED\footnote{https://ned.ipac.caltech.edu/}) so that extinction only
affects the zero points of the $PL$ relations, whereas the slope values are solid.

%-------------------------------------------------------------
%                                             Simple A&A Table
%-------------------------------------------------------------
%
\begin{table*}
\caption{Results of the least square fit in the form $mag=\alpha+\beta \log
  P$ for the LMC and SMC, where $mag$ is represented by the Wesehneit
  magnitudes $W$ (calculated as in Eq.~\ref{wdefinition}) or by the
  $G$, $G_{BP}$, $G_{RP}$ magnitudes. The different columns show: 1)
  the studied galaxy; 2) and 3) the coefficients of the linear
  regression and relative errors;  4) the r.m.s. of the residuals; 5) the number of
  objects used in the fit; 6) the method ($PL$ or $PW$); 7) the type
  of the pulsators; 8) the notes.}             % title of Table
\label{tabResults}      % is used to refer this table in the text
\centering                          % used for centering table
\begin{tabular}{c c c c c c c c}        % centered columns  (4
                                % columns)  
\hline\hline                 % inserts double horizontal lines
\noalign{\smallskip} 
 Galaxy & $\alpha$ &  $\beta$ & $\sigma$ & n & Method  & Type & Note  \\    % table heading 
(1) & (2) & (3) & (4) & (5) & (6) & (7) & (8) \\
\noalign{\smallskip} \hline \noalign{\smallskip}
LMC &   16.000$\pm$0.008 &  -3.327$\pm$0.012 & 0.104   &  1539   &      $PW$  & DCEP\_F     &     \\                                                      
LMC &   15.518$\pm$0.004 &  -3.471$\pm$0.012 & 0.087   &  1148   &  	 $PW$  & DCEP\_1O    &          \\ 
LMC &   17.326$\pm$0.014 &  -2.765$\pm$0.021 & 0.191   &  1545   &  	      $PL(G)$                  & DCEP\_F          &     \\                                                      
LMC &   16.860$\pm$0.010 &  -3.159$\pm$0.029 & 0.209   &  1158   &  	      $PL(G)$                  & DCEP\_1O    &          \\                                                      
LMC &   17.545$\pm$0.017 &  -2.580$\pm$0.025 & 0.229   &  1545   &  	      $PL(G_{BP})$           & DCEP\_F          &     \\                                                      
LMC &   17.088$\pm$0.012 &  -3.008$\pm$0.035 & 0.258   &  1176   &  	      $PL(G_{BP})$           & DCEP\_1O    &          \\                                                      
LMC &   16.859$\pm$0.012 &  -2.892$\pm$0.018 & 0.159   &  1542   &  	      $PL(G_{RP})$           & DCEP\_F          &     \\                                                      
LMC &   16.384$\pm$0.008 &  -3.204$\pm$0.023 & 0.169   &  1157   &  	      $PL(G_{RP})$           & DCEP\_1O    &          \\      
\noalign{\smallskip} \hline \noalign{\smallskip}
SMC &   16.705$\pm$0.015 &  -3.595$\pm$0.057 & 0.209   &  1126   &  	 $PW$ & DCEP\_F  & P$<$2.95 d       \\                                              
SMC &   16.608$\pm$0.021 &  -3.400$\pm$0.026 & 0.169   &   608   &  	 $PW$ & DCEP\_F  & P$\geq$2.95 d          \\                                        
SMC &   17.294$\pm$0.027 &  -2.897$\pm$0.034 & 0.219   &   613   &  	 $PW$ & DCEP\_1O &                  \\                                              
SMC &   16.823$\pm$0.008 &  -3.160$\pm$0.031 & 0.221   &  1259   &  	 $PL(G)$                 & DCEP\_F  & P$<$2.95 d       \\                                              
SMC &   16.137$\pm$0.006 &  -3.555$\pm$0.025 & 0.175   &  1226   &  	 $PL(G)$                 & DCEP\_F  & P$\geq$2.95 d    \\                                              
SMC &   17.916$\pm$0.017 &  -3.113$\pm$0.063 & 0.231   &  1110   &  	 $PL(G)$                 & DCEP\_1O &                  \\                                              
SMC &   17.722$\pm$0.030 &  -2.764$\pm$0.037 & 0.238   &   598   &  	 $PL(G_{BP})$          & DCEP\_F  & P$<$2.95 d       \\                                              
SMC &   17.274$\pm$0.009 &  -3.134$\pm$0.037 & 0.262   &  1264   &  	 $PL(G_{BP})$          & DCEP\_F  & P$\geq$2.95 d    \\                                              
SMC &   18.066$\pm$0.016 &  -2.892$\pm$0.063 & 0.229   &  1102   &  	 $PL(G_{BP})$          & DCEP\_1O &                  \\                                              
SMC &   17.891$\pm$0.035 &  -2.578$\pm$0.043 & 0.275   &   607   &  	 $PL(G_{RP})$          & DCEP\_F  & P$<$2.95 d       \\                                              
SMC &   17.431$\pm$0.010 &  -2.944$\pm$0.040 & 0.286   &  1287   &  	 $PL(G_{RP})$          & DCEP\_F  & P$\geq$2.95 d    \\                                              
SMC &   17.425$\pm$0.014 &  -3.153$\pm$0.054 & 0.201   &  1132   &  	 $PL(G_{RP})$          & DCEP\_1O &                  \\                                              
\noalign{\smallskip} \hline \noalign{\smallskip}
LMC &   16.725$\pm$0.033 &  -2.625$\pm$0.196 & 0.143   &    38   &  	 $PW$  & ACEP F     &     \\                                                      
LMC &   16.314$\pm$0.087 &  -2.564$\pm$0.506 & 0.166   &    13   &  	 $PW$  & ACEP 1O    &          \\                                                 
LMC &   17.948$\pm$0.034 &  -2.516$\pm$0.201 & 0.189   &    46   &  	 $PL(G)$                 & ACEP F           &     \\                                                 
LMC &   17.355$\pm$0.074 &  -2.749$\pm$0.455 & 0.173   &    19   &  	 $PL(G)$                 & ACEP 1O    &   \\                                                         
LMC &   18.115$\pm$0.052 &  -2.119$\pm$0.304 & 0.225   &    38   &  	 $PL(G_{BP})$           & ACEP F          &     \\                                                 
LMC &   17.561$\pm$0.111 &  -2.401$\pm$0.649 & 0.212   &    13   &  	 $PL(G_{BP})$           & ACEP 1O    &          \\                                                 
LMC &   17.496$\pm$0.036 &  -2.354$\pm$0.209 & 0.154   &    38   &  	 $PL(G_{RP})$           & ACEP F          &     \\                                                 
LMC &   16.992$\pm$0.056 &  -2.486$\pm$0.327 & 0.107   &    13   &  	 $PL(G_{RP})$           & ACEP 1O    &          \\                                                 
\noalign{\smallskip} \hline \noalign{\smallskip}
SMC &   17.185$\pm$0.042 &  -2.931$\pm$0.229 & 0.170   &    31   &  	 $PW$  & ACEP F     &     \\                                                     
SMC &   16.942$\pm$0.078 &  -3.211$\pm$0.659 & 0.213   &    13   &  	 $PW$  & ACEP 1O    &         \\                                                 
SMC &   18.380$\pm$0.041 &  -2.669$\pm$0.223 & 0.188   &    36   &  	 $PL(G)$                 & ACEP F          &     \\                                                 
SMC &   17.836$\pm$0.096 &  -2.943$\pm$0.843 & 0.284   &    15   &  	 $PL(G)$                 & ACEP 1O    &          \\                                                 
SMC &   18.569$\pm$0.052 &  -2.598$\pm$0.280 & 0.207   &    31   &  	 $PL(G_{BP})$           & ACEP F         &     \\                                                 
SMC &   17.966$\pm$0.111 &  -2.883$\pm$0.944 & 0.305   &    13   &  	 $PL(G_{BP})$           & ACEP 1O    &         \\                                                 
SMC &   17.954$\pm$0.039 &  -2.754$\pm$0.209 & 0.155   &    31   &  	 $PL(G_{RP})$           & ACEP F         &     \\                                                 
SMC &   17.489$\pm$0.088 &  -3.017$\pm$0.750 & 0.243   &    13   &  	 $PL(G_{RP})$           & ACEP 1O    &         \\                                                 
\noalign{\smallskip} \hline \noalign{\smallskip}
 LMC &   17.376$\pm$0.049 &  -2.356$\pm$0.050 & 0.162   &    80   &  	 $PW$  & T2CEP  &    \\                                                             
LMC &   18.627$\pm$0.055 &  -1.726$\pm$0.061 & 0.251   &   112   &  	 $PL(G)$                  & T2CEP  &    \\                                                             
LMC &   18.743$\pm$0.088 &  -1.484$\pm$0.093 & 0.303   &    82   &  	 $PL(G_{BP})$           & T2CEP  &    \\                                                             
LMC &   18.132$\pm$0.069 &  -1.875$\pm$0.072 & 0.238   &    83   &  	 $PL(G_{RP})$           & T2CEP  &    \\                                                             
\noalign{\smallskip} \hline \noalign{\smallskip}
SMC &   17.755$\pm$0.197 &  -2.359$\pm$0.183 & 0.233   &    15   &  	 $PW$  & T2CEP  &    \\                                                              
SMC &   19.063$\pm$0.147 &  -1.893$\pm$0.145 & 0.271   &    20   &  	 $PL(G)$                  & T2CEP  &    \\                                                              
\noalign{\smallskip} \hline \noalign{\smallskip}
\end{tabular}
\end{table*}

Operatively, to derive the relevant $PL/PW$ relationships, we adopted a standard
least-square fitting procedure with $\sigma$-clipping at
2.5-3 $\sigma$ level (typically 3 and 2.5 is used for $PW$ and $PL$,
respectively, due to the larger scatter in $PL$ relations). The number
of outliers is small because, as recalled above, the contamination of
Cepheids in LMC and SMC is very small. 
It is important to note that for the DCEP\_F in the SMC, we fitted
two different lines in different period regimes characterised by
values shorter or longer than $\sim$2.95 days. This break in the $PL/PW$
relations is well documented in the literature at all the wavelengths \citep[see
e.g.][]{Subramanian2015,Ripepi2016,Ripepi2017}. The result of the
fitting procedure is shown in Tab.~\ref{tabResults} and in Figs~\ref{plLMC}
and ~\ref{plSMC}. 
An inspection of Tab.~\ref{tabResults}  reveals that the $PL$ and
especially the $PW$ relations for the LMC are less dispersed than those for the SMC. This is
due to a depth effect generated by the well known elongation of the SMC along the
line of sight \citep[see][and references therein]{Ripepi2017}. 
We also note that the $PW$ for the LMC is much less dispersed than
the $PL$s both because the $PW$ is not affected by
reddening and because the color term in the Wesenheit magnitude takes partially into
account the intrinsic width of the instability strip. In the SMC there is less difference between the dispersion
of $PW$ and $PL$s because the dominant effect on the dispersion is the
elongation along the line of sight.  

As for the T2CEP variables, due to the paucity of the sample, to 
derive the $PL$ relations in $G$ we decided to use also objects without
the $G_{BP},G_{RP}$ magnitudes. 
After some experiments, we decided to exclude RVTAU stars from the fits because they are too
scattered and show a different slope of $PL/PW$ with respect to BLHER
and WVIR stars \citep[this effect is well documented in the
literature, see e.g.][]{Sos2008,Matsunaga2009,Matsunaga2011,Ripepi2015}.
The results of the above procedure are listed in Tab.~\ref{tabResults} and in Figs~\ref{plLMC}
and ~\ref{plSMC}. The T2CEP $PL$ relations in the SMC for $G_{BP},G_{RP}$ bands were not
calculated as the shortage of stars (only 15 usable objects) coupled with the large errors resulted
in unreliable relationships. 

We were also able to fit reasonable $PL/PW$ relations for the ACEP\_F
and ACEP\_1O variables in both the MCs. Also these results are presented in
Tab.~\ref{tabResults} and in Figs~\ref{plLMC} and ~\ref{plSMC}. 

As a final remark, we underline that the $PW$ and $PL$ relations
calculated in this paper (especially those in the $G$ band) will be
used in the {\it Cepheids\&RRLyrae SOS} pipeline \citep[][]{Clementini2018} for the Cepheid classification in the
next {\it Gaia} Data Release 3.

\section{Re-classification of \textit{Gaia} MW DR2 Cepheids}

\subsection{Comparison with the literature.}

As anticipated in the introduction, the sample of MW Cepheids presented  
in the {\it Gaia} DR2 is most likely significantly contaminated and
one of the purposes of this work is to  
clean it. 

To this aim, the first step consisted in a massive search for
alternative classification in the literature.
 The largest databases of
variable stars in the MW available are Simbad \citep{Wenger2000} and
VSX \citep[The International Variable Star
Index][https://www.aavso.org/vsx/]{Watson2006}. These sources have
been complemented/completed by several additional literature works
whose complete list is reported in the notes of
Tab.~\ref{tabLiterature}. This table reports the source
identification, equatorial coordinates and variability classification
given in {\it Gaia} DR2, as well as the literature name of the object,
the literature type(s) of variability, the period(s) and the source of
these information.  The acronyms for the variability types used in the table are listed
in Tab~\ref{tabAcronyms}.  The analysis of periods in the literature
is particularly important, as one cause of misclassification in DR2 is
the wrong period found by the {\it Cepheids\&RRLyrae SOS} pipeline, caused by the low number
of epochs available for a consistent sample of objects
\citep[in][objects with more than 12 epochs were
analysed]{Clementini2018}.

\begin{sidewaystable*}
\caption{Table with the literature classification for 1416 objects
  among the 2116 candidate Cepheids in the MW by \citet{Clementini2018}. The meaning
  of the different columns is: (1) Gaia DR2 source identification;
  (2)-(3) RA-DEC (J2000); (4) variability classification according to
  Gaia DR2; (5) name of the object in the literature;
  (6) type(s) of variability found in the literature; (7) source for
  the different type of variability, separated by a ``/'' or ``//''
  depending whether or not the period estimates of the two or more
  sources do agree; (8)-(10) period(s) present in the literature;
  (11)-(13) sources of the period(s) in columns (8)-(10). 
A glossary of 
of the variability types is reported Tab.~\ref{tabAcronyms}. The
acronyms for the literature are given in the notes of this table.
The table is published in its entirety only in the electronic edition of the journal. 
A portion including the first 15 lines is shown here for guidance regarding its form and content.}\label{tabLiterature}
\centering
\begin{tabular}{ccccccc} 
\hline\hline             
\noalign{\smallskip} 
 Source\_id & RA & DEC &  DR2 Class.  &  Lit. Name  &  Lit.  Class.  &   \\  
         &  deg & deg & & & \\ 
	(1) &  	(2) &  	(3) &  	(4) &  	(5) &  	(6)     \\ 
\noalign{\smallskip} \hline \noalign{\smallskip}
 2947530506428832768 & 101.64608 & -14.92456 & DCEP\_1O & ASAS  J064635-1455.5                 & DSCT/SXPHE &  \\
  208360790657462144  & 79.45432 & 44.47322 & WVIR & ASASSN-V J051749.04+442823.6        & RRAB & \\
  3315820030750497536 & 89.13874 & 1.70906 & DCEP\_1O & CRTS J055633.2+014232               & RRd & \\
  4044404165342126848 & 276.27275 & -34.44904 & DCEP\_MULTI & V3276 Sgr                           & RRAB  &\\
  4071594911751759872 & 280.72338 & -28.62898 & DCEP\_1O & [CAG2000] vs1f408                   & RRC  &\\
  3045809872243862400 & 106.25278 & -11.81071 & DCEP\_1O & GDS\_J0705006-114838                 & VAR  &\\
  4122020821451345664 & 259.72364 & -19.99288 & DCEP\_1O & V1836 Oph                           & RR & \\
  4051686608879712640 & 277.11621 & -27.4369 & DCEP\_1O & MACHO 175.30920.52                  & RRAB  &\\
  3099348185775497728 & 101.98331 & -7.41384 & BLHER & ASASSN-V J064755.99-072449.9        & RRAB &\\
  4114405122877842688 & 257.82295 & -23.27459 & DCEP\_MULTI & 140805                              & RRAB & \\
  4077490291331290368 & 278.75571 & -24.01232 & DCEP\_1O & 124321                              & RRC & \\
  4043821561616680448 & 270.95217 & -31.78211 & RVTAU & OGLE-BLG-RRLYR-12209                & RRAB  &\\
  4627678075752483584 & 65.96439 & -76.91188 & BLHER & OGLE-GAL-ACEP-006                   & ACEP/DCEP\_F &\\
  4072780464535420160 & 279.21002 & -26.99927 & DCEP\_F & [CAG2000] vs11f595                  & RRAB  &\\
  4594729766718946304 & 265.52038 & 27.75574 & DCEP\_1O & 28630                               & RRAB & \\
\noalign{\smallskip} \hline \noalign{\smallskip}
\multicolumn{7}{c}{Continuation} \\
\noalign{\smallskip} \hline \noalign{\smallskip}
 Lit. source & P1  & P2  &  P3  & P1 source   & P2 source & P3 source \\
&days & days & days &  &  &  \\
(7)  &	(8) &  	(9) &  	(10) &  	(11) &  	(12) & (13)  \\
\noalign{\smallskip} \hline \noalign{\smallskip}
 VSX/ASAS\_RICHARDS & 0.0954 & --- & --- & VSX/ASAS\_RICHARDS & --- & ---\\
  ASAS-SN & 0.1849 & --- & --- & ASAS-SN & --- & ---\\
  CRTS & 0.2731 & --- & --- & CRTS & --- & ---\\
 VSX & 0.3565 & --- & --- & VSX & --- & ---\\
  VSX & 0.3920 & --- & --- & VSX & --- & ---\\
  VSX/PS1 & 0.4276 & --- & --- & PS1 & --- & ---\\
 VSX & 0.4342 & --- & --- & VSX & --- & ---\\
 VSX & 0.4675 & --- & --- & VSX & --- & ---\\
  ASAS-SN & 0.4820 & --- & --- & ASAS-SN & --- & ---\\
  PS1 & 0.4838 & --- & --- & PS1 & --- & ---\\
  PS1 & 0.4895 & --- & --- & PS1 & --- & ---\\
  OGLE & 0.4959 & --- & --- & OGLE & --- & ---\\
 OGLE/VSX & 0.5042 & 1.8836 & --- & OGLE & VSX & ---\\
  VSX & 0.5048 & --- & --- & VSX & --- & ---\\
  PS1 & 0.5049 & --- & --- & PS1 & --- & ---\\
\noalign{\smallskip} \hline \noalign{\smallskip}
\end{tabular}
\tablebib{
ASAS3 \citep[All Sky Automated Survey,][]{Pojmanski1997,Pojmanski2002};  
ASAS\_RICHARDS \citep[All Sky Automated Survey re-classification,][]{Richards2012};  
ASAS-SN \citep[All-Sky Automated Survey for Supernovae,][]{Jayasinghe2018};
%ATLAS\citep[Asteroid Terrestrial-Impact Last Alert Survey,][]{Tonry2018}; 
B15 \citep{Berdnikov2015}; C01 \citep{Clement2001}; CRTS \citep[Catalina Real-Time Transient
Survey,][]{Drake2014,Drake2017}; GCVS \citep[General Catalogue of
Variable Stars,][]{Samus2017}; DR1 \citep{Clementini2016}; EROS2\_KIM
\citep[][]{Kim2014}; Hip11 \citep[][]{Dubath2011}; IOMC \citep[Integral Optical Monitoring
Camera,][]{Alfonso2012}; Kep11 \citep[][]{Debosscher2011}; LINEAR \citep[Lincoln Near-Earth Asteroid Research,][]{Palaversa2013}; NSVS
\citep[Northern Sky Variable Survey,][]{Wozniak2004,Hoffman2009}; 
OGLE \citep[Optical Gravitational Lensing
Experiment,][]{Sos2015a,Sos2015b,Sos2016,Sos2017a,Sos2017b,Sos2018}; 
PS1 \citep[Panoramic Survey Telescope \& Rapid Response System,][]{Sesar2017};
SDSS \citep{Ivezic2007};
Simbad \citep[][]{Wenger2000};
VSX \citep[The International Variable Star Index,][]{Watson2006} 
}
\end{sidewaystable*}

Among the 2116 candidate Cepheids in the MW, 1416 have some mention in
the literature. About 1008 of them have been classified in at least
one of the Cepheid sub-classes, whereas 50 objects have a generic
classification as ``variables''. The rest of the sample is composed by
a disparate collection of variability types (see
Tab.~\ref{tabLiterature} and Tab~\ref{tabAcronyms}), even if a
significant portion is represented by 121 variables classified as RR
Lyrae. As expected, the Cepheid sample in the MW from {\it Gaia} DR2 is actually contaminated
by different variability types. The literature classification is also
useful as a base for the specific re-classification that is argument
of the next section.

\begin{table*}
\caption{Results of the least square fit in the form of Eq.~\ref{eqABL} or Eq.~\ref{eqABLZ} for 
 the full MW DCEPs and T2CEPs sample (top part of the table) and for the selected sample of MW DCEPs with a full characterisation
 in terms of reddening and metallicity estimates (half-bottom part of
 the table).  The
 different columns show: 1-3) the coefficients of the non-linear fit
 and the relative errors;  4) the r.m.s. of the residuals of the ABL function; 5) the number of
  objects used in the fit; 6) the method ($PL$ or $PW$); 7) the type
  of the pulsators. To remark the differences with
  Tab.~\ref{tabResults}, we added an underscript ``A'' or a
  superscript 0 to show that the magnitudes adopted are absolute and/or
  de-reddened, respectively.
"Fixed" means that the slope of LMC ($\beta$) was imposed to derive the other parameters;
$PW_AZ$ indicates a PW relation depending on metallicity.}             % title of Table 
\label{resultsMW}      % is used to refer this table in the text 
\centering                          % used for centering table 
\begin{tabular}{ c c c c c c c }        % centered columns (4 columns) 
\hline\hline                 % inserts double horizontal lines 
\noalign{\smallskip} 
 $\alpha$ &  $\beta$ & $\gamma$ & $\sigma_{\rm ABL}$ & n & Method  & Type\\    % table heading 
(1) & (2) & (3) & (4) & (5) & (6) & (7) \\
\noalign{\smallskip}
\hline 
\noalign{\smallskip}
\multicolumn{7}{c}{Full MW DCEPs and T2CEPs sample} \\
\noalign{\smallskip}
\hline 
\noalign{\smallskip}  
  -2.701$\pm$0.086  &  -3.320$\pm$0.107    &                &     0.013     &    489   	  &    $P W_A$  & DCEP\_F   \\  
  -2.976$\pm$0.131  &  -4.095$\pm$0.304    &                &     0.020     &    138   	  &    $P W_A$  & DCEP\_1O   \\ 
  -1.194$\pm$0.061  &  -2.381$\pm$0.080    &                &     0.071     &    269   	  &    $P W_A$  & T2CEP \\  
\noalign{\smallskip}
\hline 
\noalign{\smallskip}
  -2.699$\pm$0.023  &  -3.327  fixed       &                 &     0.013   &    489    	  &    $P W_A$  & DCEP\_F   \\     
  -3.246$\pm$0.045  &  -3.471  fixed       &                 &     0.020   &    138    	  &    $P W_A$  & DCEP\_1O   \\    
  -1.211$\pm$0.043  &  -2.356  fixed       &                 &     0.071   &    269    	  &    $P W_A$  & T2CEP   \\ 
\noalign{\smallskip}
\hline 
\noalign{\smallskip}\multicolumn{7}{c}{Selected MW DCEPs sample} \\
\noalign{\smallskip}
\hline 
\noalign{\smallskip}  
  -2.837$\pm$0.081  &  -3.183$\pm$0.097    &              &     0.011        &    292    &     $P W_A$  & DCEP\_F         \\ 
  -3.214$\pm$0.223  &  -3.587$\pm$0.507    &              &     0.012        &     33    &     $P W_A$  & DCEP\_1O         \\
  -1.942$\pm$0.096  &  -2.454$\pm$0.116    &              &     0.025        &    273    &     $PL(G_A^0)$             & DCEP\_F      \\  
  -1.903$\pm$0.302  &  -3.709$\pm$0.712    &              &     0.026        &     33    &     $PL(G_A^0)$             & DCEP\_1O      \\
  -1.816$\pm$0.102  &  -2.229$\pm$0.121    &              &     0.031        &    273    &     $PL(G_{BP,A}^0)$           & DCEP\_F       \\       
  -2.100$\pm$0.178  &  -2.776$\pm$0.375    &              &     0.022        &     33           &     $PL(G_{BP,A}^0)$           & DCEP\_1O       \\
  -2.313$\pm$0.094  &  -2.607$\pm$0.113    &              &     0.019        &    273    &     $PL(G_{RP,A}^0)$           & DCEP\_F        \\
  -2.637$\pm$0.178  &  -3.110$\pm$0.383    &              &     0.016        &     33    &     $PL(G_{RP,A}^0)$           & DCEP\_1O       \\
\noalign{\smallskip} \hline \noalign{\smallskip}
  -2.721$\pm$0.025  & -3.327  fixed       &                 &  0.011      &    292   &      $P W_A$  & DCEP\_F         \\ 
  -3.261$\pm$0.056  & -3.471  fixed       &                 &  0.012      &     33   &      $P W_A$  & DCEP\_1O         \\
  -1.688$\pm$0.032  & -2.765  fixed       &                 &  0.025      &    273   &      $PL(G_A^0)$                  & DCEP\_F       \\ 
  -2.175$\pm$0.072  & -3.159  fixed       &                 &  0.028      &     33   &      $PL(G_A^0)$                  & DCEP\_1O      \\
  -1.525$\pm$0.033  & -2.580  fixed       &                 &  0.031      &    273    &      $PL(G_{BP,A}^0)$           & DCEP\_F        \\
  -1.855$\pm$0.040  & -3.308  fixed       &                 &  0.023      &     33   &      $PL(G_{BP,A}^0)$           & DCEP\_1O       \\
  -2.083$\pm$0.030  & -2.892  fixed       &                 &  0.019      &    273   &      $PL(G_{RP,A}^0)$           & DCEP\_F           \\   
  -2.593$\pm$0.042  & -3.204  fixed       &                 &  0.016      &     33   &      $PL(G_{RP,A}^0)$           & DCEP\_1O       \\
\noalign{\smallskip}
\hline 
\noalign{\smallskip}
  -2.862$\pm$0.082  &  -3.134$\pm$0.095    &  -0.237$\pm$0.199  &     0.011&    261    	  &      $PW_AZ$    & DCEP\_F         \\ 
\noalign{\smallskip}
\hline 
\noalign{\smallskip}
  -2.716$\pm$0.028  &  -3.327  fixed       &  -0.105$\pm$0.207  &     0.011&    261    	  &      $PW_AZ$    & DCEP\_F         \\ 
\noalign{\smallskip}
\hline 
\noalign{\smallskip}
\end{tabular}
\end{table*}

%-------------------------------------- Two column figure (place early!) 
   \begin{figure*}
   \centering 
\vbox{
   \includegraphics[width=9cm]{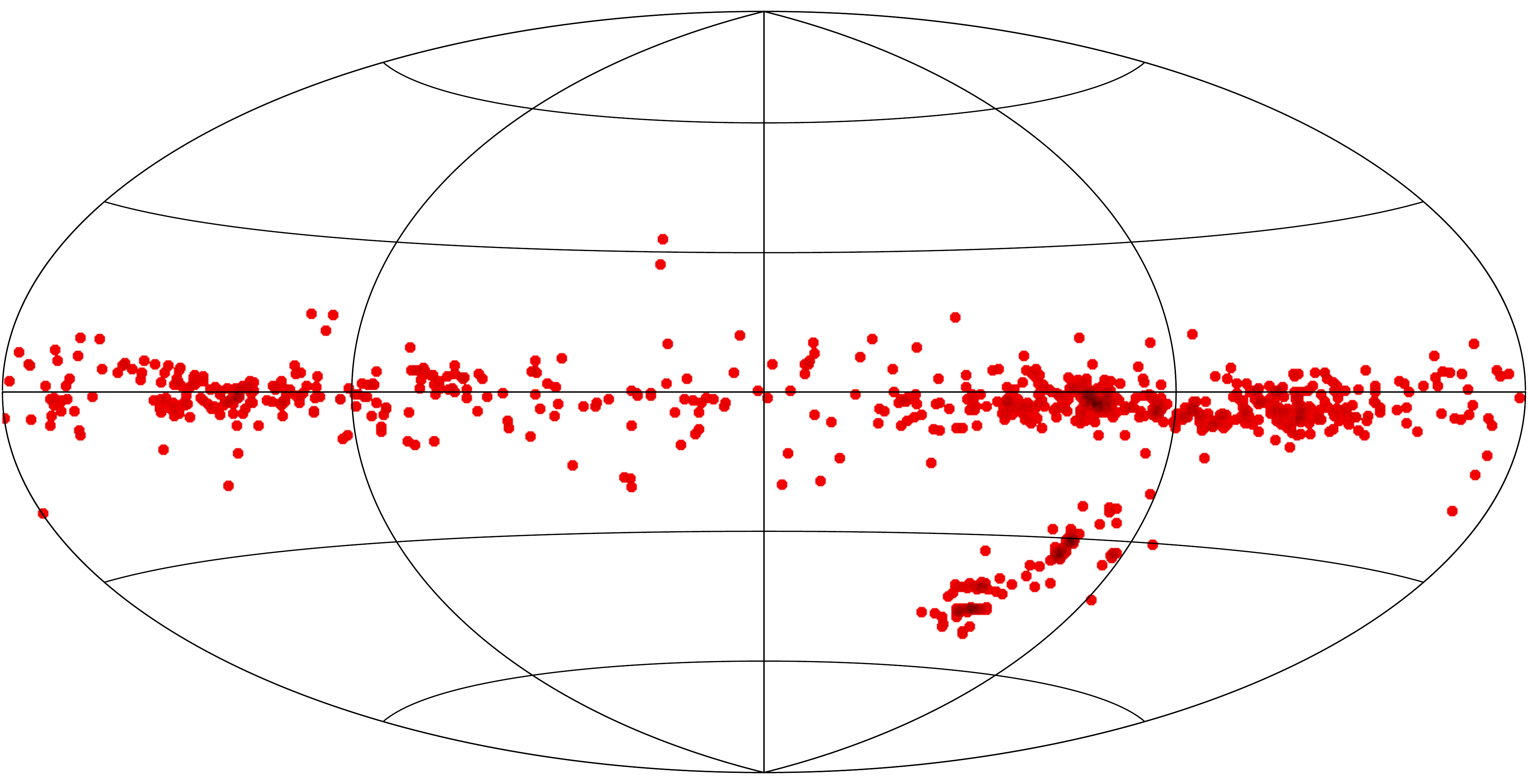}
   \includegraphics[width=9cm]{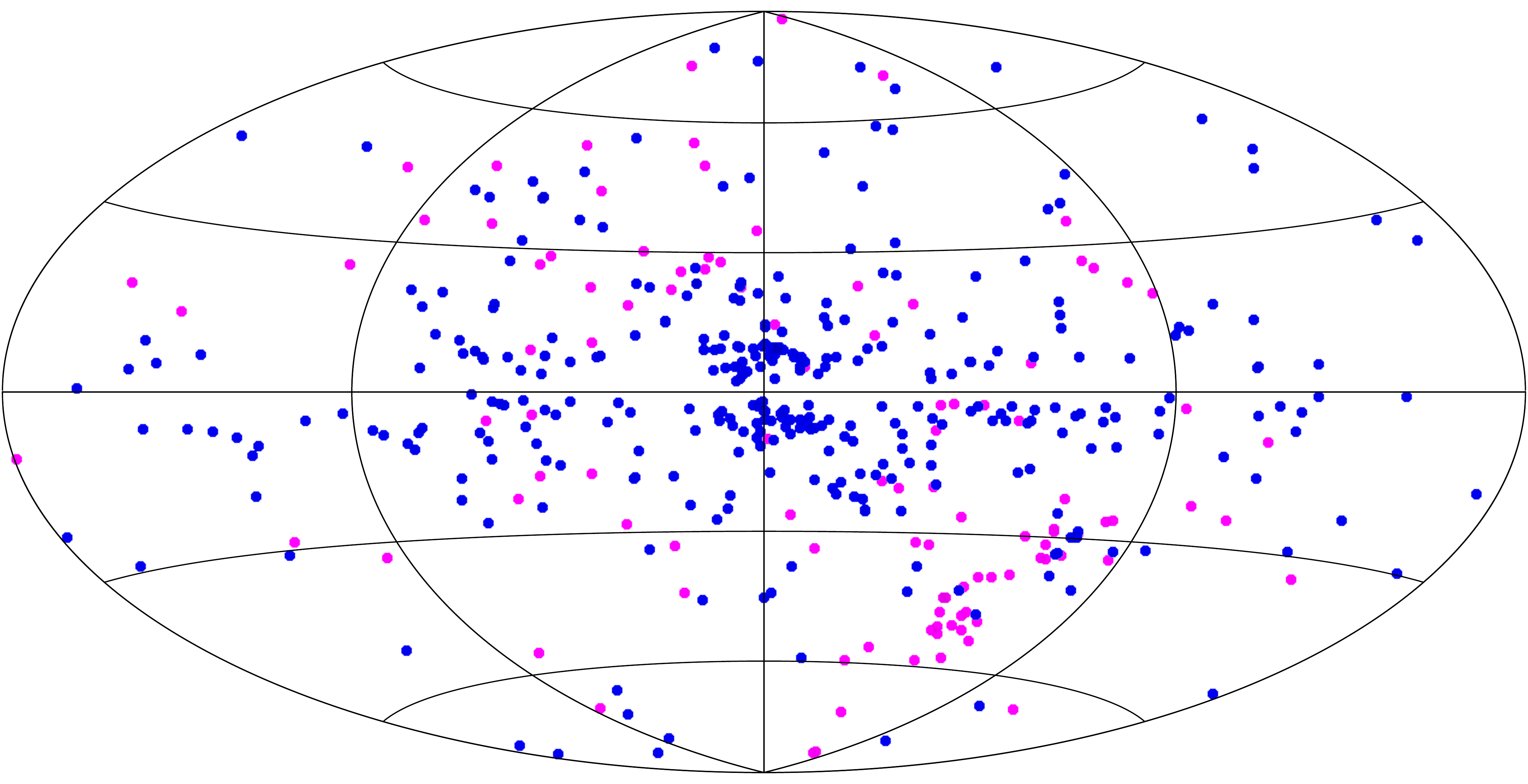}
}
   \caption{Aitoff projection in galactic coordinates of the 
     objects re-classified in this work. Left panel: DCEPs (pulsating 
     in any mode - red filled circles). Right panel: T2CEPs (of any type - blue filled circles) and ACEPs 
     (pulsating in any mode - magenta filled circles).}
              \label{figSky}%
    \end{figure*}

%-------------------------------------- Two column figure (place early!) 
   \begin{figure}
   \centering 
\vbox{
   \includegraphics[width=9cm]{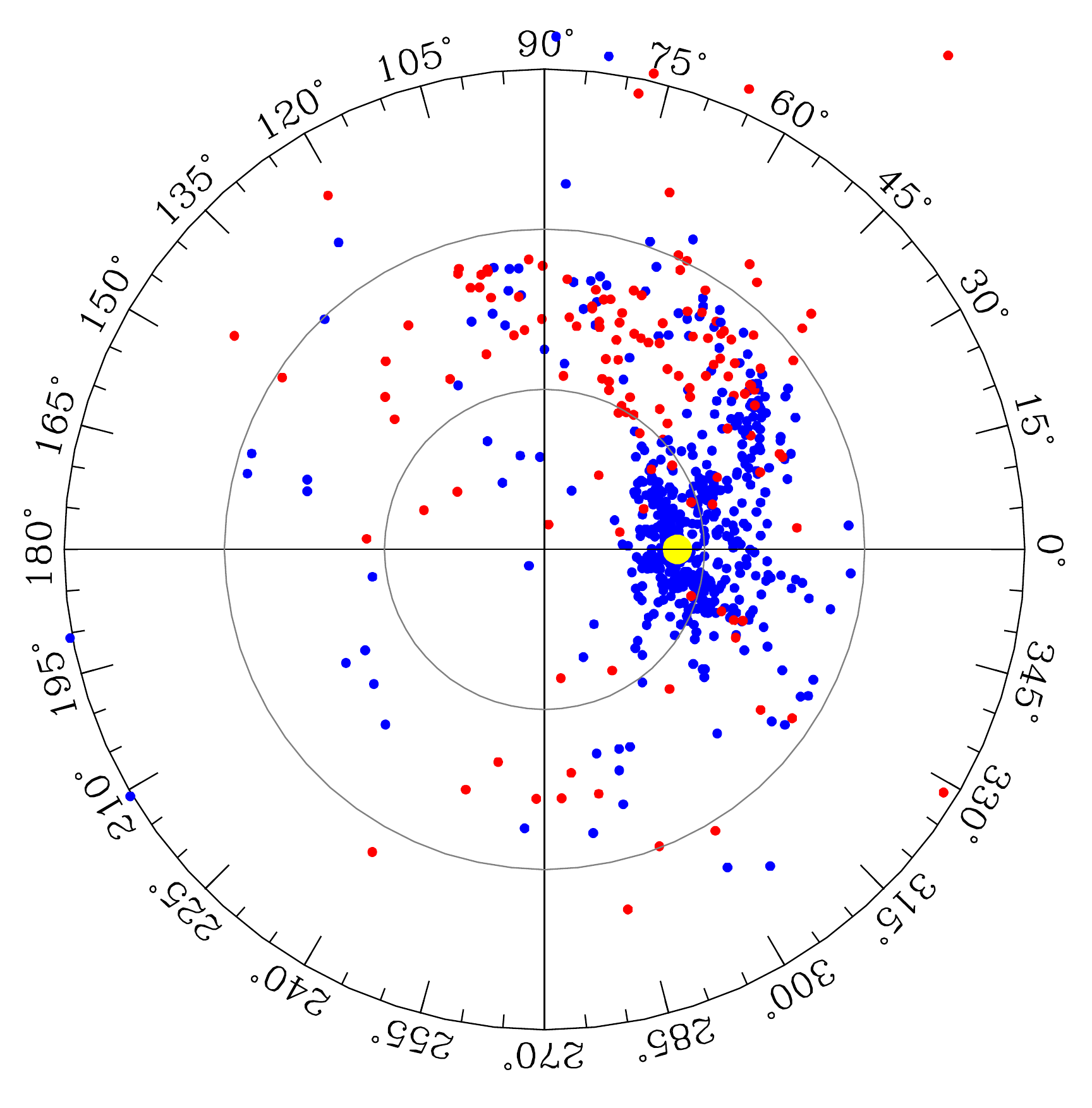}
   \includegraphics[width=9cm]{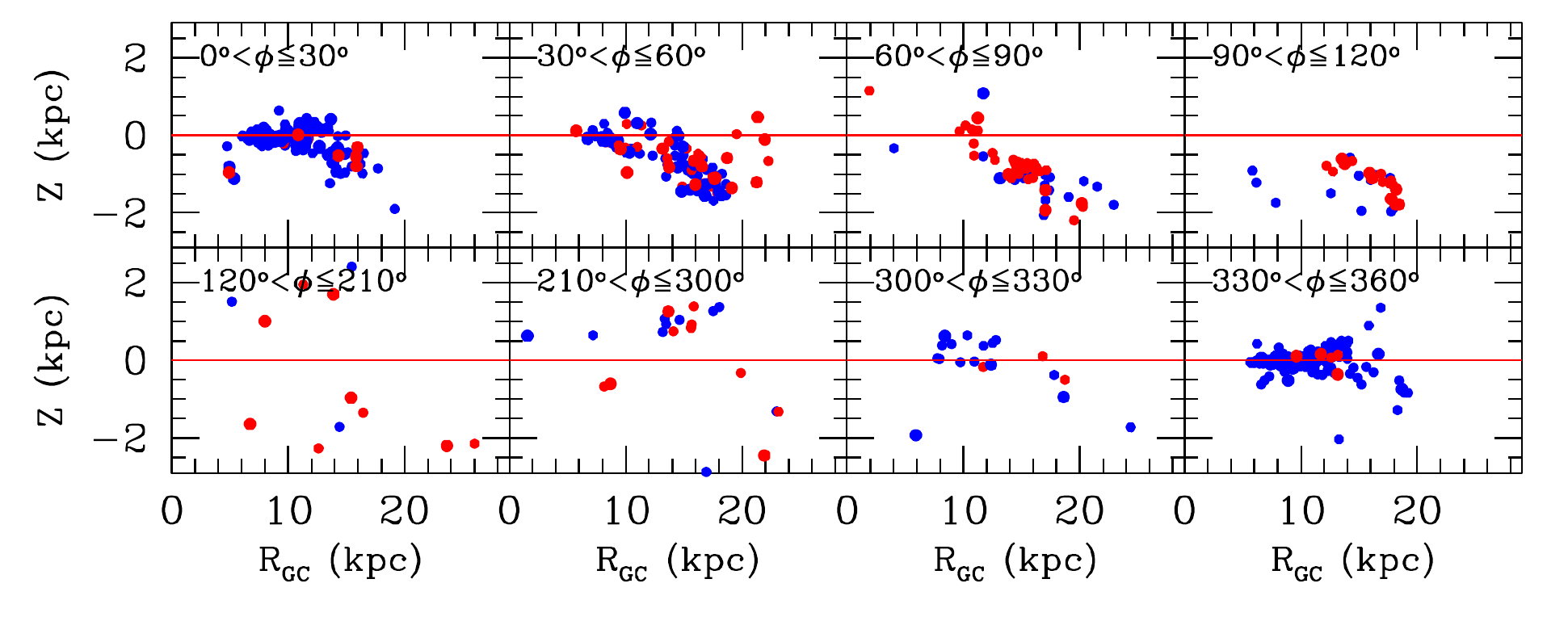}
}
   \caption{Top panel: polar map of the galactic plane depicted by known (blue filled circles)
     and newly discovered (red filled circles) DCEPs (pulsating in any
     mode). The galactic center is in the middle, the circles have radii of 10, 20 and 30
     kpc respectively. A yellow disk represents the position of the
     Sun. Note that the Galactocentric polar coordinate $\Phi$ is 0$\degr$ in
     the direction of the Sun.
     Bottom panel: distribution of the distances from the
     galactic plane (Z) as a function of the Galactocentric distance
     (R$_{\rm GC}$) for different intervals of $\Phi$. The warping of the disk is clearly visible.}
              \label{figGalaxy}%
    \end{figure}

\subsection{Detailed re-classification}

The procedure adopted for the re-classification relies on the visual
inspection of each light curve (LC) and comparison with a reliable
atlas of LCs such as that by the OGLE
group\footnote{http://ogle.astrouw.edu.pl/atlas/index.html} for the
classical pulsating stars. The visual inspection of LCs was
complemented by the analysis of the location of the stars in
period-absolute Wesenheit magnitude \citep[or Astrometry-Based
Luminosity, ABL, in case of negative parallaxes, see e.g.][and
Eq.~\ref{eqABL} in the next section]{Arenou1999} and period-Fourier
parameters (R$_{21}$, R$_{31}$, $\phi_{21}$, $\phi_{31}$) diagrams
\citep[for a definition of the Fourier parameters and their use in the
{\it Gaia Cepheids\&RRLyrae SOS} pipeline, see][and references
therein]{Clementini2016,Clementini2018}. Additionally, we took into
account the peak-to-peak amplitude ratio Amp($G_{BP}$)/Amp($G_{RP}$)
that, for the different types of Cepheids assumes characteristic
values as shown in Tab.~\ref{tabAmpRatio}. The amplitude ratio is
particularly useful to separate non-pulsating from pulsating
variables, as the former assumes generally values smaller
($\sim$1.0-1.2) than the latter ($\sim$1.3-1.6).

\begin{table}
\caption{Average peak-to-peak amplitude ratio Amp($G_{BP}$)/Amp($G_{RP}$)
for the different types of Cepheids considered in this work. Note that
the value listed in the table have been calculated on the
re-classified sample.}             % title of Table
\label{tabAmpRatio}      % is used to refer this table in the text
\centering                          % used for centering table
\begin{tabular}{l c c }        % centered columns (4 columns)
\hline\hline                 % inserts double horizontal lines
\noalign{\smallskip} 
 Type & Amp. Ratio  & dispersion \\
\noalign{\smallskip} \hline \noalign{\smallskip}
 DCEP\_F       &   1.58  &   0.10   \\
DCEP\_1O    &   1.63  &   0.07 \\
ACEP           &  1.54   &  0.20   \\
BLHER         &   1.53  &   0.14   \\
WVIR           &  1.33   &  0.14   \\
RVTAU        &  1.45   &  0.25  \\
\noalign{\smallskip} \hline \noalign{\smallskip}
\end{tabular}
\end{table}

We note that in building the $PW/P-$ABL diagrams we have corrected the
parallax zero point by adding 0.046 mas according to
\citet{Riess2018b} \citep[see also][for a recent discussion on the zero
point offset]{Schonrich2019}.
This operation has little importance for the
purpose of re-classifying the Cepheids, but is important for the
determination of the absolute $PW$ relations performed in the next
section. The re-classification made also use of the literature
classification, that was especially useful in the most doubtful
cases. In particular, the use of periods from the literature helped to
reclassify more than 140 objects whose LC shape revealed clearly wrong
period from {\it Gaia} DR2, generally caused by the low number of
epochs available for these objects. In several of these cases, when
  sufficient data were available, we used the {\it Gaia} photometry 
to recalculate the periods using the literature values as starting
point and refining them using {\it Period04} package \citep{Lenz2005}.

In this context, particularly useful was the work by \citet{Lemasle2018},
  who analysed in detail the multimode DCEPs in DR2, providing a list
  of reliable new multimode candidates.

As for the types of variability, our re-classification is restricted
to all subtypes of Cepheid variables that are the main target of
present work. Apart from these objects, we only classified in detail
RR Lyrae and ACEP stars. The former because their characteristic LCs make them
relatively easy to identify and because they are analysed together
with Cepheids in the {\it Gaia Cepheids\&RRLyrae SOS} pipeline.
As for the ACEPs, they were absent in the  {\it Cepheids\&RRLyrae SOS} pipeline
\citep[see][]{Clementini2018}, but several literature works
reported their presence in the MW (e.g. OGLE group). Moreover during
the process of re-classification, we realised that the LC shape for
some short (1-2 days) period Cepheids did not fit completely neither with DCEPs nor
with BLHERs. Therefore, we adopted the usual classification scheme for ACEPs
in terms of ACEP\_F and ACEP\_1O. Note that in the absence of very
precise distances (the candidate ACEPs are in general faint), the distinction 
between ACEP\_1O and DCEP\_1O on the basis of the LC shapes
is very difficult, because at fixed period the LCs of these two
classes are very similar. Similarly, ACEP\_Fs with periods
shorter/longer  than one day can be confused with RRABs or DCEP\_Fs,
respectively. The distinction between ACEPs and DCEPs is favoured by
the position of the object in the MW (high galactic latitude DCEPs
are unlikely), whereas RRABs are ubiquitous in the Galaxy, and a similar 
separation cannot be carried out. 
The distinction between these classes will be greatly facilitated by
the availability of more precise parallaxes,
as expected from the next {\it Gaia} data releases.

In the end, the classification types considered here are:
DCEP\_F, DCEP\_1O, DCEP\_2O, DCEP\_MULTI\footnote{DCEP\_MULTI class is 
 in turn subdivided into subclasses according to the period ratios of the modes pulsating
  simultaneously, for example fundamental/first overtone (F/1O). In
  this paper when we classify an object as DCEP\_MULTI, we are assuming
  that the period ratios found in the {\it
  Gaia} DR2 is correct.}, ACEP\_F, ACEP\_1O, BLHER, WVIR,
RVTAU, CEP, RRAB, RRC, OTHER, where CEP means that the object is a
Cepheid candidate but we could not determine the type.

Before proceeding with the analysis (i.e. the construction of $PW/P-$ABL
diagrams), we have checked the goodness of the {\it Gaia} astrometric
solution for the 2116 MW DR2 Cepheids. According to
\citet{Lindegren2018} a parameter measuring the goodness of the fit is
the {\it astrometric\_excess\_noise} ($\epsilon_i$), measuring the
excess of noise of the source. If $\epsilon_i>$0, the residuals are
statistically larger than expected. The additional parameter {\it
  astrometric\_excess\_noise\_sig} ($D$) measures the significance of
$\epsilon_i$. If $D\leq$2 then $\epsilon_i$ is probably not significant,
and the source could have a good astrometric solution even if
$\epsilon$i is large. More recently, \citet{Lindegren2018tn} devised a
new parameter called RUWE (re-normalised unit weight error), not part
of the official {\it Gaia} DR2, that consists in a renormalisation of
the astrometric ${\chi}^2$. According to \citet{Lindegren2018tn}
values of RUWE $\leq$1.4 should indicate good astrometry. We
cross-matched the two indicators and decided to take as objects with
good astrometry those with RUWE $\leq$1.4, $\epsilon_i\leq$1,
$D\leq$2, resulting in 151 out of  2116 stars with not reliable
astrometry. The position of these stars in the $PW/P-$ABL relations was
not taken into account for the classification, which then was based
only on the shape of the LCs and on the Fourier parameters.

\begin{sidewaystable*}
\caption{Table with the new classification. The meaning of the different columns is: (1) Gaia DR2 source identification; (2)-(3) RA-DEC (J2000); (4)-(6) number of epochs in $G$, $G_{BP}$ and $G_{RP}$, respectively; (7) Period; (8)-(10)
intensity averaged magnitudes in $G$, $G_{BP}$ and $G_{RP}$, respectively; (11)-(12) and (13)-(14) Fourier amplitude ratios and Fourier phase differences; (15)-(16) parallax and parallax error; (17) excess of flux in the BP and RP integrated
photometry with respect to the G band; (18)-(19) excess noise of the source and its significance; (20) re-normalised unit weight error (see text); (21) mode of pulsation from present work;
(22) flag "N" to denote a Cepheid not known in the literature; (23) Notes (see the end of the table for details). Note that when a numeric value is missing we assigned the value ``99.999'', whereas in case 
of empty string fields, we  display a ``---'' string. 
The table is published in its entirety only in the electronic edition of the journal. 
A portion including the first 15 lines is shown here for guidance regarding its form and content.}\label{tabReclassified}
\centering
\begin{tabular}{cccccccccccc} 
\hline\hline             
\noalign{\smallskip} 
 Source\_id & RA & DEC & n$_{G}$  &  n$_{G_{BP}}$  & n$_{G_{RP}}$  & Period & $G$  &  $G_{BP}$  & $G_{RP}$  & $R_{21}$ & $R_{31}$ \\
         &  deg & deg & & & & days & mag & mag & mag &  &  \\
	(1) &  	(2) &  	(3) &  	(4) &  	(5) &  	(6) &  	(7) &  	(8) &  	(9) &  	(10) &  	(11) &  	(12) \\
\noalign{\smallskip} \hline \noalign{\smallskip}
 2947530506428832768 & 101.64608 & -14.92456 & 19 & 17 & 17 & 0.09537 & 12.677 & 12.997 & 12.22 & 99.999 & 99.999\\
  3315820030750497536 & 89.13874 & 1.70906 & 14 & 14 & 14 & 0.27332 & 16.263 & 17.001 & 15.385 & 0.086 & 0.131\\
  4071594911751759872 & 280.72338 & -28.62898 & 14 & 14 & 14 & 0.39195 & 16.182 & 16.4 & 15.775 & 0.186 & 99.999\\
  4122020821451345664 & 259.72364 & -19.99288 & 20 & 16 & 16 & 0.43408 & 16.396 & 16.761 & 15.785 & 0.287 & 99.999\\
  4296338318281270144 & 293.05243 & 8.93232 & 32 & 28 & 32 & 0.44584 & 16.755 & 17.361 & 15.992 & 0.136 & 0.133\\
  6641655551574932992 & 298.6506 & -53.31556 & 32 & 31 & 31 & 0.45379 & 13.83 & 14.016 & 13.499 & 0.13 & 99.999\\
  5631368811358200320 & 141.58416 & -31.61573 & 19 & 16 & 17 & 0.46519 & 16.371 & 16.604 & 15.981 & 0.229 & 99.999\\
  4051686608879712640 & 277.11621 & -27.4369 & 17 & 16 & 16 & 0.46752 & 15.764 & 16.085 & 15.262 & 99.999 & 99.999\\
  4437777711669788416 & 244.96412 & 5.29721 & 39 & 33 & 35 & 0.46976 & 16.403 & 16.598 & 16.067 & 0.078 & 0.074\\
  4113412916685330304 & 254.11532 & -24.06865 & 36 & 30 & 28 & 0.47605 & 16.5 & 16.869 & 15.891 & 0.07 & 0.082\\
  4311050922079027584 & 285.55926 & 10.57317 & 18 & 15 & 13 & 0.47706 & 18.337 & 19.361 & 17.269 & 0.158 & 99.999\\
  3099348185775497728 & 101.98331 & -7.41384 & 19 & 17 & 16 & 0.48199 & 14.245 & 14.649 & 13.646 & 99.999 & 99.999\\
  4114405122877842688 & 257.82295 & -23.27459 & 21 & 17 & 19 & 0.48384 & 16.982 & 17.432 & 16.387 & 0.186 & 0.425\\
  5958267083020200448 & 263.74736 & -44.83491 & 22 & 21 & 22 & 0.48582 & 15.313 & 15.668 & 14.777 & 0.126 & 99.999\\
  1454878497455250048 & 205.5597 & 28.42065 & 45 & 44 & 42 & 0.48612 & 15.315 & 15.451 & 14.968 & 0.09 & 0.11\\
\noalign{\smallskip} \hline \noalign{\smallskip}
\multicolumn{12}{c}{Continuation} \\
\noalign{\smallskip} \hline \noalign{\smallskip}
$\phi_{21}$ & $\phi_{31}$ & $\varpi$ & $\sigma_\varpi$ & E(BR/RP) &  $\epsilon_i$ &  $D$  &  RUWE &  Mode &  New &  Notes &\\
rad & rad & mas & mas &  & mas &  & & & & & \\
(13) &  (14) &  (15) &  (16) &  (17) &  (18) &  (19) &  (20) &  (21) & 	(22) &  	(23) & \\
\noalign{\smallskip} \hline \noalign{\smallskip}
 99.999 & 99.999 & 0.855 & 0.03 & 1.205 & 0.0 & 0.0 & 0.965 & OTHER & --- & a\\
  0.838 & 0.395 & 0.415 & 0.078 & 1.344 & 0.0 & 0.0 & 1.013 & RRC & --- & ---\\
  2.439 & 99.999 & -0.096 & 0.089 & 1.217 & 0.091 & 0.355 & 1.044 & RRC & --- & a\\
  3.251 & 99.999 & 0.068 & 0.077 & 1.301 & 0.0 & 0.0 & 0.999 & RRAB & --- & a,e\\
  1.688 & 2.749 & 0.123 & 0.076 & 1.279 & 0.0 & 0.0 & 1.032 & OTHER & --- & ---\\
  2.243 & 99.999 & 0.177 & 0.026 & 1.199 & 0.108 & 3.37 & 1.228 & RRC & N & f,g\\
  2.012 & 99.999 & 0.018 & 0.084 & 1.245 & 0.167 & 1.133 & 1.079 & OTHER & --- & ---\\
  99.999 & 99.999 & 0.401 & 0.103 & 1.274 & 0.45 & 10.49 & 1.599 & RRAB & --- & a\\
  2.393 & 5.797 & 0.046 & 0.066 & 1.208 & 0.0 & 0.0 & 0.985 & OTHER & --- & ---\\
  1.695 & 5.894 & 0.107 & 0.086 & 1.247 & 0.0 & 0.0 & 1.003 & OTHER & --- & ---\\
  2.451 & 99.999 & 0.064 & 0.207 & 1.447 & 0.55 & 1.565 & 1.051 & --- & --- & NC:c,e\\
  99.999 & 99.999 & 0.229 & 0.032 & 1.237 & 0.0 & 0.0 & 0.911 & RRAB & --- & a\\
  1.914 & 1.335 & -0.127 & 0.128 & 1.311 & 0.0 & 0.0 & 1.023 & RRAB & --- & ---\\
  1.898 & 99.999 & 0.04 & 0.041 & 1.234 & 0.0 & 0.0 & 1.065 & DCEP\_1O & N & ---\\
  2.299 & 4.807 & -0.045 & 0.035 & 1.234 & 0.0 & 0.0 & 0.993 & RRC & --- & h\\
\noalign{\smallskip} \hline \noalign{\smallskip}
\end{tabular}
\tablefoot{Notes: a) wrong period in DR2, used the literature one or derived in this work; b) uncertain period; c) uncertain astrometry; d) incomplete light curve; e) scattered light curve; f) bad astrometric solution (see text);
g) classification based on the light curve shape; h) adopted literature classification; i) uncertain classification of light curve shape. }
\end{sidewaystable*}

Having set out all the tools, we proceeded with the re-classification
by looking first at the position of the star in the $PW/P-$ABL
relations. Due to the large relative error on parallax, the position
of the targets in these diagrams is often ambiguous, i.e. compatible
with different Cepheid types. This occurs in particular for periods
shorter than 3 days, characteristics of DCEPs, ACEPs and
BLHERs. Moreover, DCEPs and WVIRs candidate positions largely overlap
when the relative error on the parallax is larger than $\sim$30\%.  We
then passed to a visual inspection of the LCs and of the
period-Fourier parameters diagrams. Particularly useful were the
P-R$_{21}$ and the P-$\phi_{21}$ diagrams to separate DCEP\_F from
DCEP\_1O and low-period DCEP\_F from ACEP\_F and BLHER,
respectively. Despite all these efforts, in some cases, the
classification of Cepheids with saw-tooth LC shape and periods
$\sim1-2$ days was difficult, as the shape of the LCs of DCEP\_F,
ACEP\_F and BLHER are very similar in this period range and the
differences can only be revealed in very well sampled and precise LCs,
a condition not always fulfilled in our case. Also the position of
these objects in the P-$\phi_{21}$ diagram was sometimes not
conclusive.  In some ambiguous cases we assigned to the ACEP class 
 objects with high galactic latitude, as we do not expect DCEPs in the
 MW halo. A similar distinction cannot be carried out between BLHER
 and ACEPs, as these classes share the same locations in the MW.
In any case, the classification of these objects might be
subject to a revision when more accurate {\it Gaia} parallaxes (as
well as metallicity estimates, given that both ACEPs and BLHERs are
expected to be more metal poor with respect to DCEPs) will be
available and will allow us to disentangle clearly the $PL/PW$ relations
for the different Cepheid types as it happens in the LMC/SMC. For 13
objects with clear Cepheid-like LC and correct position in the
P-Fourier parameters diagrams we were not able to assign a more
precise type, and we indicated them with CEP. Their detailed
sub-classification will be determined using future {\it Gaia} releases.

\begin{sidewaystable*}
\caption{Comparison of the re-classified object with the
 literature. Columns and rows show the classification given in this
  work and in the literature, respectively. The ``NEW''
  and ``TOTAL'' rows show the number of new objects found in this work
and the total number for each pulsating class.}             % title of Table 
\label{tabReclassifiedvsLit}      % is used to refer this table in the text 
\centering                          % used for centering table 
\begin{tabular}{|c|c|c|c|c|c|c|c|c|c|c|c|c|c|c|c|}
\noalign{\smallskip} \hline
% \multicolumn{1}{|c|}{col1} &
% \multicolumn{1}{c|}{col2} &
% \multicolumn{1}{c|}{col3} &
% \multicolumn{1}{c|}{col4} &
% \multicolumn{1}{c|}{col5} &
% \multicolumn{1}{c|}{col6} &
% \multicolumn{1}{c|}{col7} &
% \multicolumn{1}{c|}{col8} &
% \multicolumn{1}{c|}{col9} &
% \multicolumn{1}{c|}{col10} &
% \multicolumn{1}{c|}{col11} &
% \multicolumn{1}{c|}{col12} &
% \multicolumn{1}{c|}{col14} &
% \multicolumn{1}{c|}{col13} &
% \multicolumn{1}{c|}{col15} \\
%\hline
   & DCEP\_F & DCEP\_1O & DCEP\_2O & DCEP\_M & ACEP\_F & ACEP\_1O &
                                                                    BLHER
                            & WVIR & RVTAU & RRAB & RRC & OTHER & & CEP & NC\\
\hline 
  DCEP\_F & 449 & 6 & 0 & 1 & 5 & 0 & 16 & 19 & 1 & 0 & 0 & 14 & &2 & 7\\
\hline 
  DCEP\_1O & 7 & 129 & 0 & 0 & 0 & 1 & 4 & 2 & 0 & 0 & 0 & 6 & &0 & 0\\
\hline 
  DCEP\_2O & 0 & 0 & 1 & 0 & 0 & 0 & 0 & 0 & 0 & 0 & 0 & 0 & &0 & 0\\
\hline 
  DCEP\_M & 0 & 0 & 0 & 16 & 0 & 0 & 0 & 0 & 0 & 0 & 0 & 1 & &0 & 1\\
\hline 
  ACEP\_F & 7 & 0 & 0 & 0 & 54 & 1 & 5 & 0 & 0 & 0 & 0 & 0 & &0 & 0\\
\hline 
  ACEP\_1O & 0 & 0 & 0 & 0 & 0 & 2 & 0 & 0 & 0 & 0 & 0 & 0 & &0 & 0\\
\hline 
  BLHER & 10 & 5 & 0 & 0 & 3 & 1 & 56 & 0 & 0 & 0 & 0 & 3 & &4 & 0\\
\hline 
  WVIR & 4 & 0 & 0 & 0 & 0 & 0 & 0 & 89 & 5 & 0 & 0 & 11 & &1 & 8\\
\hline 
  RVTAU & 0 & 0 & 0 & 0 & 0 & 0 & 0 & 2 & 22 & 0 & 0 & 7 & &0 & 16\\
\hline 
  RRAB & 1 & 0 & 0 & 0 & 18 & 0 & 8 & 0 & 0 & 79 & 0 & 0 & &0 & 5\\
\hline 
  RRC & 0 & 1 & 0 & 0 & 0 & 0 & 0 & 0 & 0 & 0 & 2 & 0 & &0 & 1\\
\hline 
  OTHER & 29 & 12 & 0 & 0 & 1 & 0 & 4 & 10 & 19 & 0 & 1 & 179 & &0 & 42\\
\hline 
&&&&&&&&&&&&&&& \\
\hline 
  NEW & 68 & 51 & 0 & 3 & 21 & 1 & 49 & 24 & 1 & 1 & 1 & 426 & &6 & 86\\
\hline 
&&&&&&&&&&&&&& &\\
\hline 
  TOTAL & 575 & 204 & 1 & 20 & 102 & 6 & 142 & 146 & 48 & 80 & 4 & 647 & &13 & 128\\
\hline \noalign{\smallskip}
\end{tabular}
\end{sidewaystable*}

The result of the procedure described above is shown in Table~\ref{tabReclassified}, where
we report for each of the 2116 MW Cepheid the new classification as
well as all the data from {\it Gaia} DR2 used in the re-classification
process. These include the parameters to estimate the goodness of the
astrometry and the parameter E(BP/RP), indicating the excess of flux
in the  $G_{BP},G_{RP}$ bands with respect to the $G$ band. Values
larger than 2 usually indicates problems with colors. This parameter
is reported for completeness but it affects just very few objects. 
A detailed description of the
different columns can be found in the table caption. In the notes
(last column) we report special cases, e.g. when the literature
period was used, the astrometry not usable etc. An inspection of the
table reveals that no classification was
possible for 128 objects, due to various reasons specified in the
notes, being the most common ones the lack of precise parallaxes
and/or scanty/incomplete LCs.

  \begin{figure}
   \centering 
   \includegraphics[width=9cm]{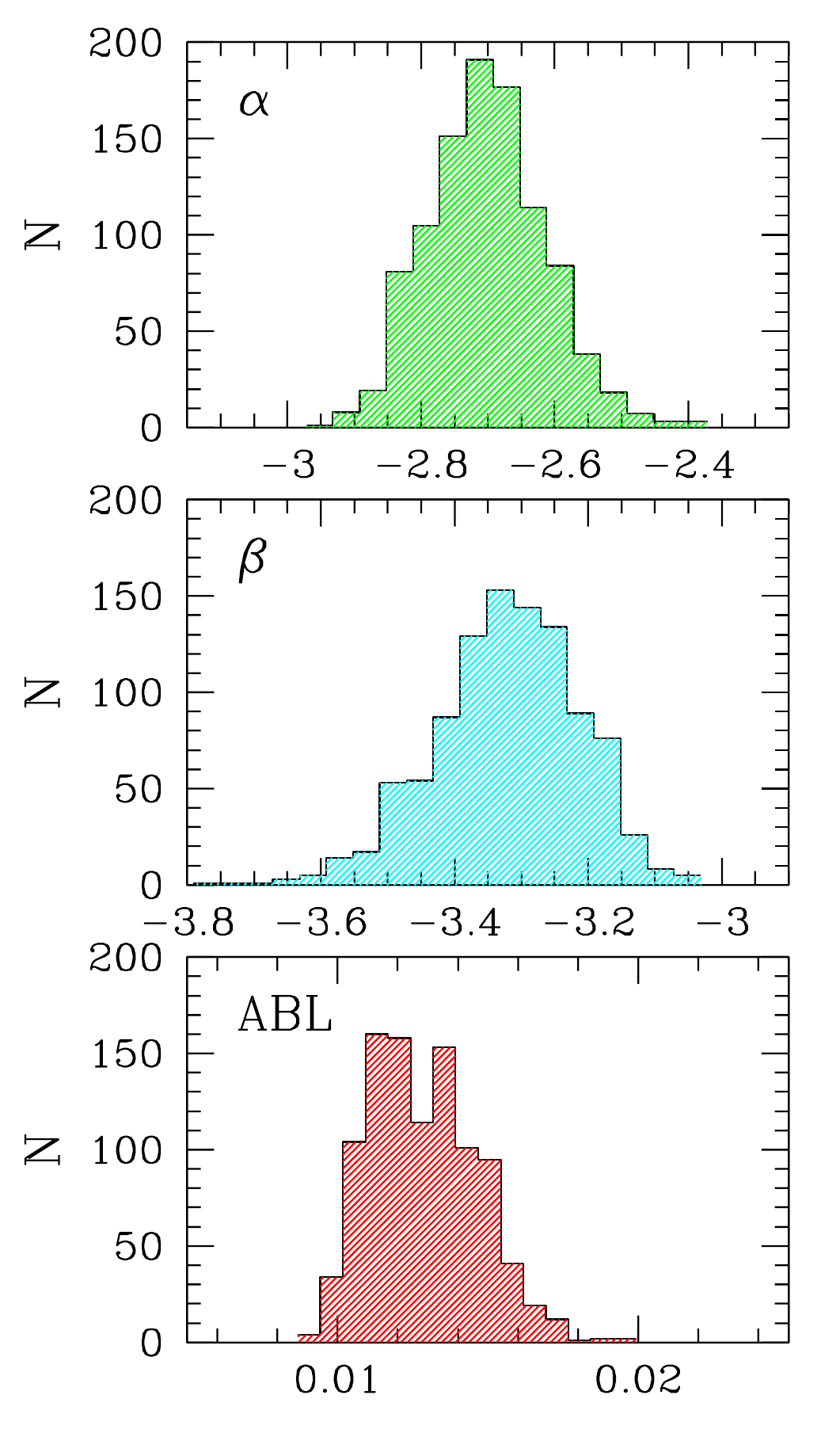}
   \caption{Example of the results of the bootstrap procedure 
     described in the text in the case of the $PW$ in the form $W_A$=$\alpha+\beta\log P$
     applied to DCEP\_F. From top to bottom, the different panels show 
     the distribution of the parameters $\alpha$, $\beta$ and of the 
     residuals of data around the  ABL function, respectively. 
}              \label{figBootstrap}%
\end{figure}

A total of 1257 stars have been classified as Cepheid of any type, 
84 objects as RR Lyrae and 647 as variables of other type (in addition
to the 128 stars with no classification).

An overall comparison of the new classification for the 1257 Cepheids with the
literature is shown in Table~\ref{tabReclassifiedvsLit}. An
inspection of the table shows that we have changed the literature
classification for 270 objects, whereas 274 are new Cepheids
completely unknown in the literature or indicated as ``variable''.

To visualize the results, we show in Fig.~\ref{figPLMW} (upper panels)
the $PW$ relations for the stars classified as Cepheids except those
ones with negative parallaxes  (184 objects). Errorbars are not shown
for clarity reasons. 
The different types of variables are identified in the figure with different
colors (see caption of the figure). An inspection of the figure reveals that due to the large
errors in parallaxes, objects belonging to different Cepheids types
are mixed and it is not easy to define tight $PW$ relations as those
for the LMC/SMC. The situation is improved if we restrict to objects
with relative error on parallaxes lower than 20\%. This is shown in
the lower panels of  Fig.~\ref{figPLMW}. 

We can compare these results with those reported in Fig. 7 of
\citet{Clementini2018}. A large part of the objects below
the dashed line in that figure, more than 700 objects that were expected to be contaminating
stars, now disappear and are classified as ``OTHER'' or not
classified (about 150 of them were known in the literature as non pulsating
variables, see Table~\ref{tabLiterature}).   
However not all the objects in the lower part of the diagram
disappeared, as several objects that are clearly Cepheid variables can be found
several magnitudes below (some also above) the relevant PW
sequence. This is not surprising since, among the other issues: i) the astrometric solution for DR2 
did not take into account duplicity and therefore the presence of
companions can affect not only the photometry, but also the parallax;
ii) the chromatic correction for the astrometric solution is
based on the mean magnitude and not on the epoch colour. \citep[see][]{Lindegren2018}.

The P-R$_{21}$/$\phi_{21}$ and P-R$_{31}$/$\phi_{31}$ diagrams for the
re-classified Cepheids are shown in Fig.~\ref{figFourier}. A comparison
with the similar Figs. 37 and 38 of \citet{Clementini2018} show that
the sequences of the different types of Cepheids are now better
defined and more congruent with those in the MCs.

Similarly, the location on the sky in galactic coordinates for the
re-classified Cepheids is shown in Fig. ~\ref{figSky}. Left and right panels
of the figure display the location of DCEPs and ACEPs/T2CEPs
respectively. The DCEPs are now located preferentially along the MW
disk, as expected for this {population I} stars, whereas ACEP/DCEP are
 distributed more homogeneously across the MW including the center and
 the halo, as expected 
 \citep[compare with Fig. 39 in][]{Clementini2018}.

  \begin{figure*}
   \centering 
\vbox{
   \includegraphics[width=9cm]{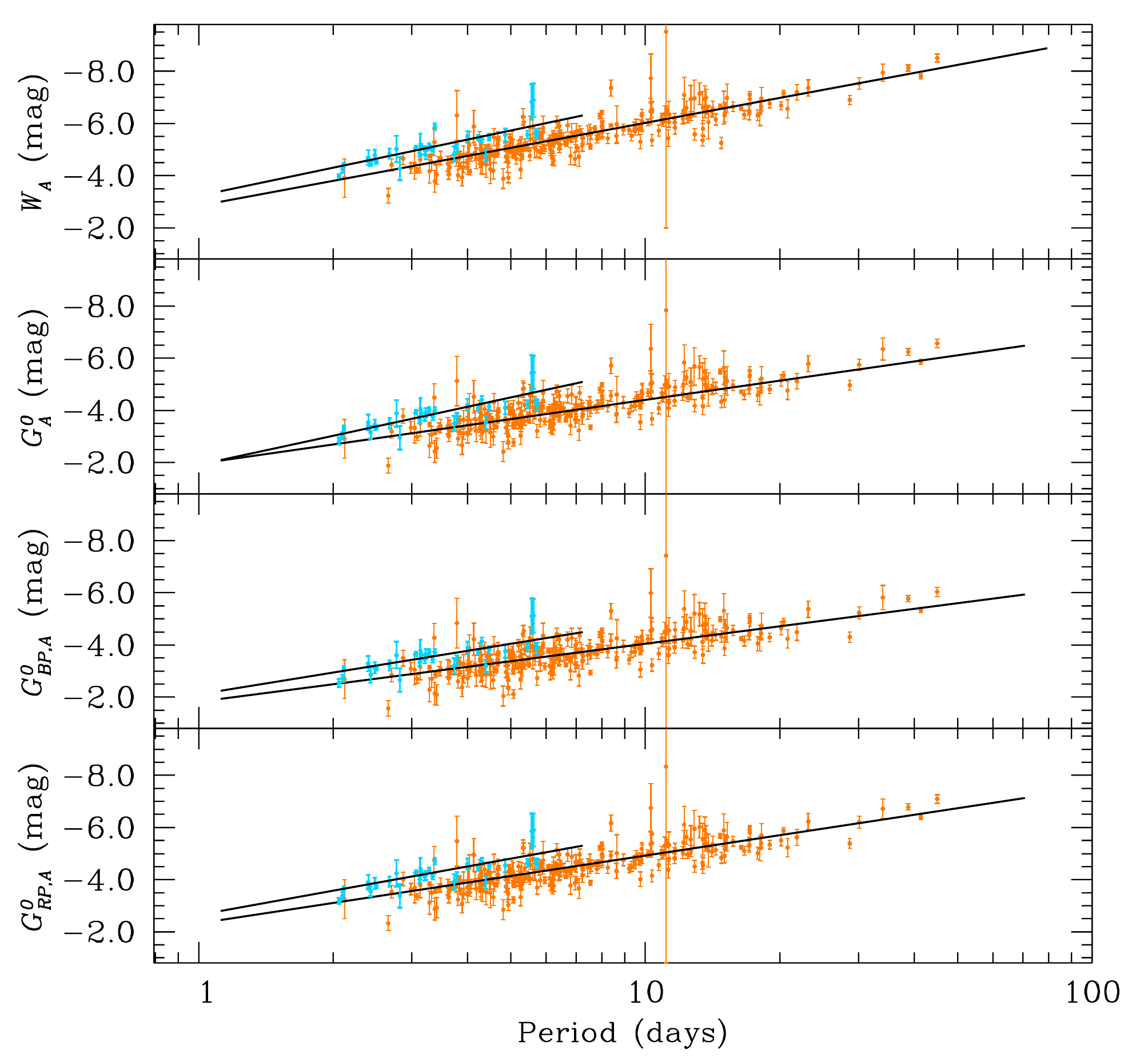}
   \includegraphics[width=9cm]{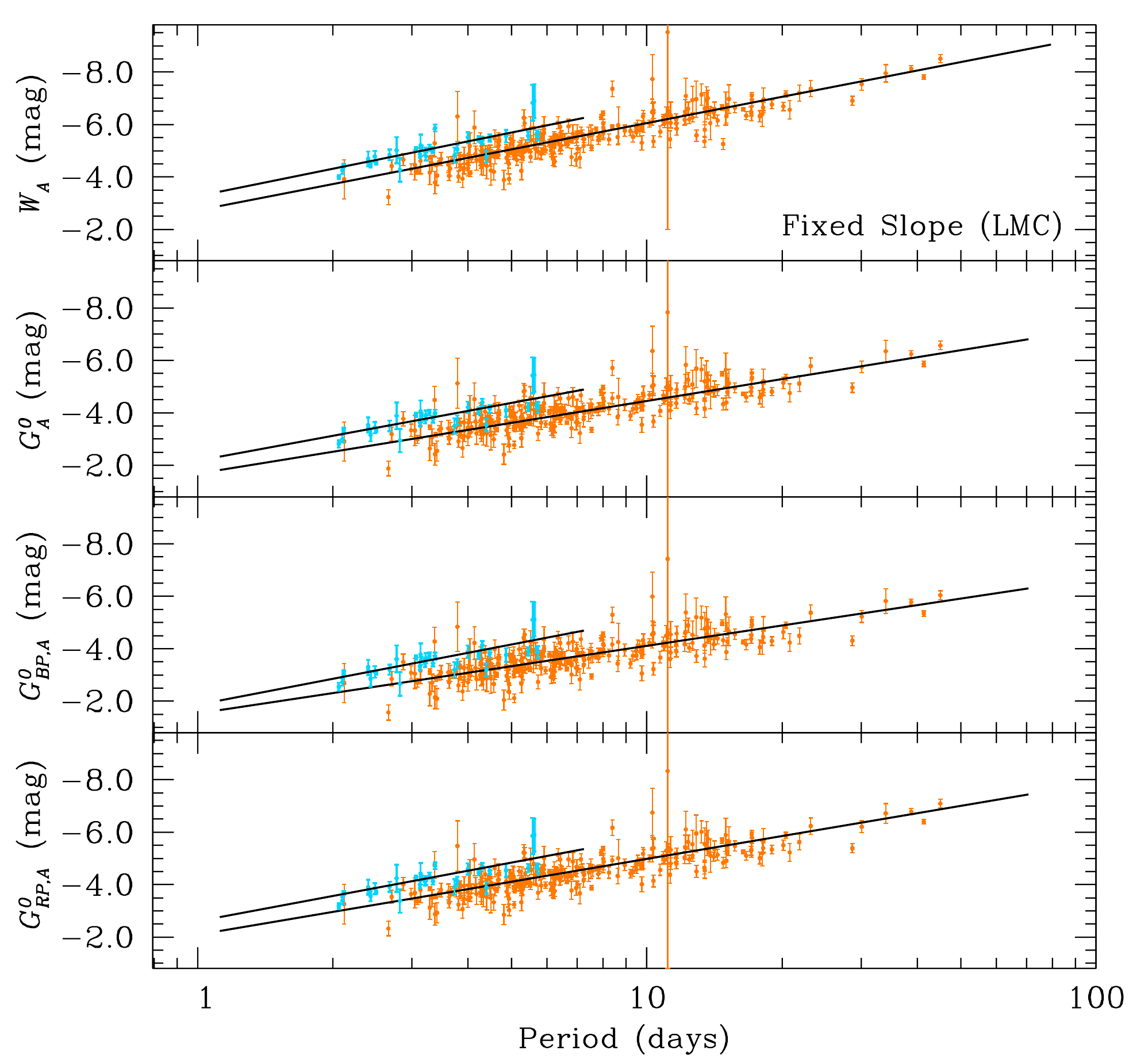}
}
   \caption{ $PL/PW$ relations for the MW selected sample of DCEPs having 
     reddening estimate and intensity averaged magnitudes in $G$, $G_{BP}$ and 
 $G_{RP}$ bands coming from the  {\it Cepheids\&RRLyrae SOS} pipeline. Orange and light blue 
symbols represent DCEP\_F and DCEP\_1O, respectively. The solid lines 
are the least-square fits to the data obtained using the ABL 
formulation (see text). As in Fig. ~\ref{figPLMW}, left and right panels 
show the relations obtained leaving all the parameters free to vary 
and fixing the value of $\beta$ in Eq.~\ref{eqABL}, respectively. 
The coefficient of the regression lines are shown in Tab.~\ref{resultsMW}
An underscript ``A'' means absolute magnitudes whereas a 
superscript 0 characterises de-reddened quantities.
}              \label{figSelected}%
\end{figure*}

  \begin{figure}
   \centering 
\vbox{
   \includegraphics[width=9cm]{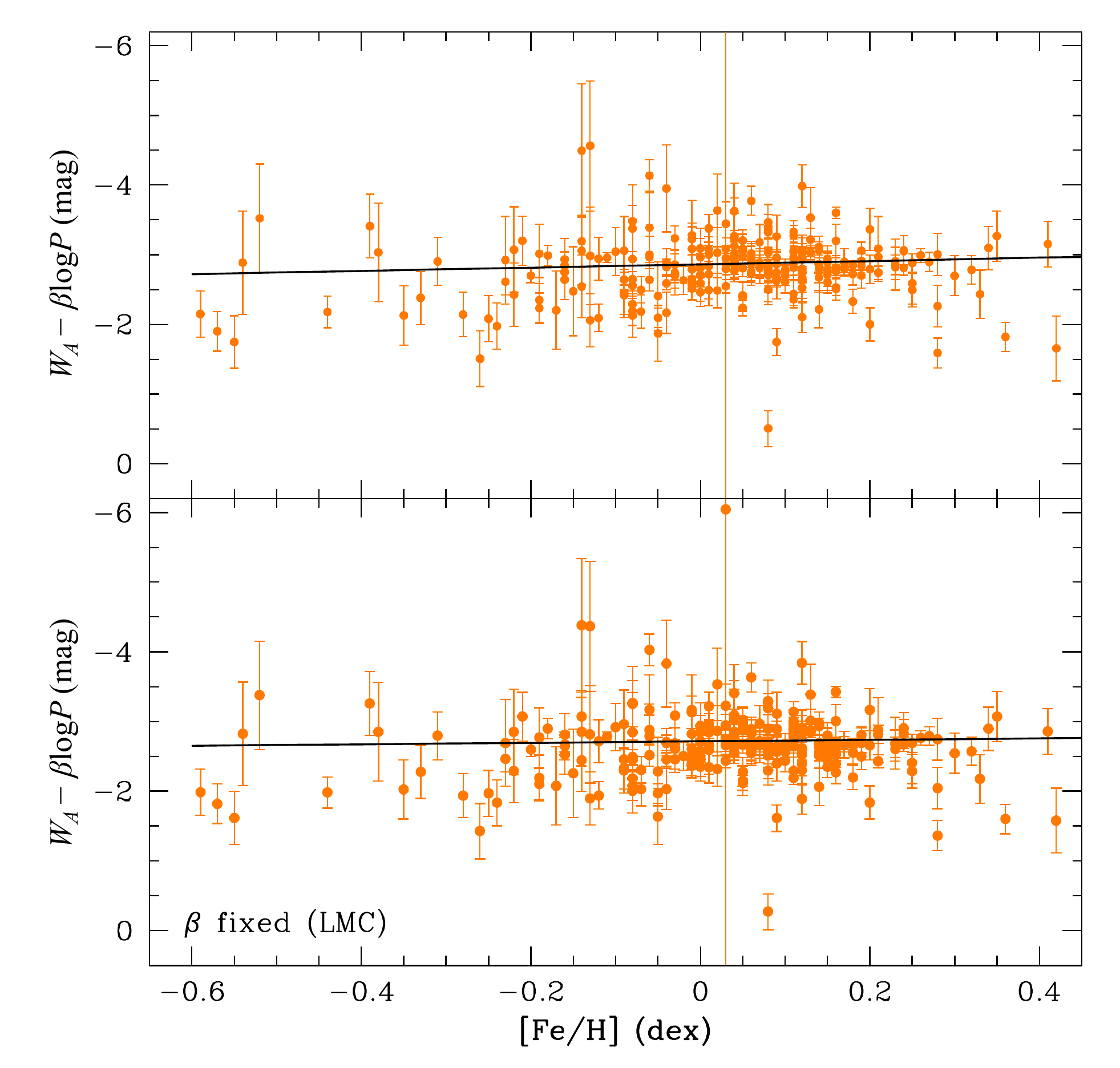}
}
   \caption{Dependence of the $PW$ relation from $[Fe/H]$. Orange 
     symbols represent DCEP\_F pulsators, whereas the solid lines 
are the results of the fitting procedure for the ABL formulation of 
Eq.~\ref{eqABLZ} in two cases: i) all parameters free to vary (top 
panel); ii) $\beta$ parameter fixed to the value of LMC (bottom panel).  
The coefficient of the regression lines are shown in the last two 
lines of Tab.~\ref{resultsMW}
}              \label{figFeh}%
\end{figure}

\subsubsection{Cepheid stars hosted by stellar clusters or dwarf 
  galaxies orbiting the  MW}

\label{altriSistemi}

Having completed the re-classification, we checked whether some of the objects comprised in the MW sample 
 is actually hosted by a stellar systems such as Galactic open or 
globular clusters (OC, GC) or by dwarf galaxies orbiting the MW. 
To reach our goal,  we i) inspected the literature and ii) tested new 
possible associations. 
As for the literature, we relied on the work by \citet{Anderson2013}
and by \citet{Clement2001}  for the association between DCEPs and open 
clusters and between RRLs/ACEPs/T2CEPs and GCs. 
Different sources were adopted for the association with 
dwarf galaxies in the local group. The result of this 
work is reported in Tab.~\ref{inAltriSistemi}. An inspection of the 
table shows that 53 and 66 Cepheids of different types were already 
known from the OGLE survey to be hosted by LMC and SMC, respectively. 

We also searched additional associations between Cepheids in the MW 
sample and the above quoted stellar systems.  However, we did not investigate new 
associations between DCEPs and OCs, as this complex work would 
deserve an entire new paper.  
We searched new MCs objects by simply overlapping the Cepheids in the 
surroundings of these galaxies (i.e. from 
-56$\degr\leq$Dec$\leq$-80$\degr$, 0h$\leq$Ra$\leq$4h and 
4h$\leq$Ra$\leq$8h for the SMC and LMC, respectively) with the precise 
$PL/PW$ relations holding for these systems. In case an object with a 
certain Cepheid type falls within 3$\sigma$ of the relative $PL/PW$ 
sequences (Tab.~\ref{tabResults}), we considered a positive match and assigned the object to 
the LMC or SMC. In this way we assigned 8 and 7 new Cepheids of 
different types to the LMC and SMC, respectively (see 
Tab.~\ref{inAltriSistemi} for details). 
Thus we have a total of 61 and 73 Cepheids hosted by the LMC and SMC,
respectively. These objects were then used to derive the $PL/PW$
relations for the MCs calculated in Sect.~\ref{sectMCs} and listed in
Tab.~\ref{tabResults}. The effect of the few tens DCEPs added to the
LMC/SMC samples is insignificant, whereas the addition of the ACEPs
increased significantly the sample.

\begin{table*}
\caption{Table with the association of pulsator in the All Sky sample
  with open/golular clusters as well as with dwarf galaxies satellites
  of the MW. The meaning
  of the different columns is: (1) Gaia DR2 source identification;
  (2) name of the object in the literature (if any);
  (3) type of variability according to this work; (4) host system; (5)
  source of the association of the variable with the stellar system. 
The table is published in its entirety only in the electronic edition of the journal. 
A portion including the first 15 lines is shown here for guidance regarding its form and content.}\label{inAltriSistemi}
\centering
\begin{tabular}{ccccc} 
\hline\hline             
\noalign{\smallskip} 
 Source\_id & Lit. Name &  DR2 Class.  &  Host system  &  Source of association. \\ 
	(1) &  	(2) &  	(3) &  	(4) &  	(5)     \\ 
\noalign{\smallskip} \hline \noalign{\smallskip}
  428620663657823232 &  DL Cas    & DCEP\_F & NGC129  & A13\\
  429385923752386944 &  CG Cas   & DCEP\_F & Berkeley58   & A13\\
  2011892320749270912 &   CE Cas B   & DCEP\_F & NGC7790  & A13\\
  2011892325047232256 &   CE Cas A   & DCEP\_F & NGC7790  & A13\\
  2011892703004353792 & CF Cas    & DCEP\_F & NGC7790  & A13\\
  2031776202613700480 &   SU Cyg    & DCEP\_F & Turner9  & A13\\
  4085919765884068736 &  BB Sgr   & DCEP\_F & Collinder394   & A13\\
  4092905375639902464 & U Sgr                & DCEP\_F & IC4725  & A13\\
  4094784475310672128 &  WZ Sgr    & DCEP\_F & Turner2   & A13\\
  4156512638614879104 &  EV Sct            & DCEP\_1O & NGC6664   & A13\\
  5835124087174043136 &   S Nor    & DCEP\_F & NGC6087  & A13\\
  5891675303053080704 &  V Cen    & DCEP\_F & NGC5662   & A13\\
  5932565900081831040 &  QZ Nor   & DCEP\_1O & NGC6067   & A13\\
  5932569709575669504 &   V340 Nor   & DCEP\_F & NGC6067  & A13\\
  2957940098405233024 & V7 & WVIR & NGC1904 &  C01\\
\noalign{\smallskip} \hline \noalign{\smallskip}
\end{tabular}
\tablebib{
A13 \citep{Anderson2013};
B05 \citep{Bersier2002};
C01 \citep{Clement2001}; 
CO15 \citep{Coppola2015};
DR1 \citep{Clementini2016}; 
EROS2\_KIM \citep[][]{Kim2014};
K08  \citep{Kinemuchi2008}; 
MV16 \citep{MartinezVasquez2016};
OGLE \citep[Optical Gravitational Lensing
Experiment,][]{Sos2015a,Sos2015b,Sos2016,Sos2017a,Sos2017b,Sos2018}; 
TW (This Work).
}
\end{table*}

As for the possible association with GCs or 
other dwarf galaxies in the local group, we cross-matched the position 
of the Cepheids in the  MW sample with the positions of these objects, 
looking for objects within the tidal radii of GCs or within twice the 
semimajor axes of the  dwarf galaxies \citep[we adopted][for the positions and the cluster tidal 
radii/dwarf galaxies semi-major axes values, respectively]{Harris1996,McConnachie2012}. 
We then used {\it Gaia} DR2 photometry and proper motions (PMs) to check if 
the target has a position in the Color-Magnitude diagram (CMD) and PMs 
compatible with the rest of the stars of the investigated system.  
As a result of this exercise, we were able to associate 1 ACEP\_F
variable with  the URSA MINOR dwarf spheroidal galaxy, 1 WVIR pulsator 
with the GC NGC 6254 and a variable of unknown type to 
NGC 6266 (see Tab.~\ref{inAltriSistemi}).

\subsubsection{Distribution of the MW DCEPs on the galactic plane.}

 To further show the properties of the clean DCEPs sample, it is
  interesting to investigate the distribution of these pulsators on 
  the galactic plane. To this aim, we first calculate the Galactocentric cartesian distances by subtracting the
heliocentric space vector of the Galactic centre, $\overrightarrow{D}{_0}$ from
the heliocentric space vector of our targets $\overrightarrow{D}{_{\odot}}$:
 
\begin{equation}
\overrightarrow{D}{_{\rm GC}}= \overrightarrow{D}{_{\odot}}-\overrightarrow{D}{_0} 
\end{equation}

\noindent or

\begin{equation}
\begin{pmatrix} X \\ Y \\ Z \end{pmatrix}=\begin{pmatrix} d\cos(b)\cos(l) \\ d\cos(b)\sin(l) \\
dsin(b) \end{pmatrix}-\begin{pmatrix} D_0 \\ 0 \\ 0 \end{pmatrix}
\end{equation}

\noindent
with $D_0$ being the distance of the Sun from the Galactic centre and
$l$, $b$ and $d$ the Galactic longitude, Galactic latitude and heliocentric
distance, respectively, of each DCEP.  The heliocentric distances $d$
in kpc were obtained from the $PW$ obtained for the MW DCEP\_F sample
(first line of Tab.~\ref{resultsMW}, see next section) using the simple equation:

\begin{equation}
d=10^{0.2(W-W_A)-2} 
\end{equation}

\noindent 
where $W$ and $W_A$ are the apparent and absolute Wesenheit
magnitudes, respectively.  We used the same procedure also for 
DCEP\_1O (because their $PW$ relation is much more uncertain), by
fundamentalising their periods using the equation $P_F =
P_{1O}/(0.716-0.027 \log P_{1O})$, being $P_F$  and $P_{1O}$ the
periods of DCEP\_F and DCEP\_1O, respectively \citep[][]{Feast1997}.   
Finally, the distance of the targets from the Galactic centre is given as:

\begin{equation}
{\rm R_{GC}}=\sqrt{[d \cos(b) \cos(l) - D_0]^2+d^2 \cos(b)^2 \sin(l)^2+d^2 \sin(b)^2}
\end{equation}

The distribution of DCEPs on the galactic plane is shown in the top
panel of Fig.~\ref{figGalaxy}, where known and newly
discovered DCEPs are depicted with blue and red symbols,
respectively. The figure shows that, as expected, most of the known pulsators are
placed within few kpc from the Sun, whereas the majority of the new
ones are further away. Note also that the DCEPs investigated here
trace the Local Arm as well as the Perseus Arm.  It is also
interesting to look at the distribution of the pulsators around the
galactic plane. This is displayed in the bottom panel of Fig.~\ref{figGalaxy},
where we plot the height (Z) of each object as a function of the
Galactocentric distance ${\rm R_{GC}}$ for selected intervals of the
Galactocentric angular coordinate $\Phi$ that is 0 in the
direction of the Sun and increases counterclockwise.
The figure shows clearly the presence of the well known disk warp,
especially for $0 \degr<\Phi<120 \degr$. These results are in
agreement with the works by \citet{Chen2019,Skowron2018}, 
who used different DCEP samples to study the warp of the MW disk. 
A detailed discussion of the warp as traced by DCEPs,  is beyond
the scope of present paper and we remand the interested reader to the
quoted papers for in depth discussions on the arguments.

\section{$PW$ relations for MW Cepheids}
\label{sectPW}

The new dataset of re-classified Cepheids allows us to derive the $PW$
relation directly from the data for DCEPs and T2CEPs. We preferred not
to try with ACEPs, due to the paucity of the sample and the
considerable dispersion in the $PW$ plane, resulting from the large parallax errors
(ACEPs are generally significantly fainter than DCEPs). 
Note that the 107 DCEPs belonging to  
LMC/SMC (see Sect.~\ref{inAltriSistemi})  were excluded from the  MW
DCEP sample adopted for the following analysis to avoid contamination
by much less metallic objects with respect to the MW ones. We decided
not to exclude T2CEPs from both MCs and other GC/Dwarf galaxies
satellites of the MW because the properties of these objects are
expected to be more homogeneous in different environments.

To use all the Cepheids in our sample we are forced to adopt only the Wesenheit
magnitude, as we do not know individual reddenings for each Cepheid,
making impossible for the moment to derive meaningful $PL$ relations.
Similarly, we did not attempt to add a metallicity term in
Eq.~\ref{eqABL} (see below), as this information is lacking for a consistent part
of our sample.

To derive the $PW$ relations we decided to use the ABL defined
below. We underline that the adoption of this quantity has the
decisive advantage to use the parallax in a linear fashion, avoiding 
almost any kind of bias, as no selection is done on the Cepheid
sample. Indeed, the employment  of the ABL allows us to 
include in the analysis objects with negative parallaxes. A detailed discussion of
the advantages of the ABL method is present in other papers \citep[see
e.g. ][]{Arenou1999,GaiaClementini2017} to which we refer the
interested readers.  

The ABL for the absolute Wesenheit magnitude $W_A$ is defined as follows:

\begin{equation}
{\rm ABL}=10^{0.2 W_A}=10^{0.2(\alpha+\beta\log P) }=\varpi10^{0.2{W}-2} 
\label{eqABL}
\end{equation}

\noindent
where we used the definition of $PW$ relation: $W_A$=$\alpha+\beta\log P$;
 $W_A$ and $W$ are the absolute and relative Wesenheit
magnitudes, respectively. The observed quantities are $W$, $P$ and $\varpi$. 
The unknown $\alpha$ and $\beta$ values can be obtained using a
least-square fit procedure.   

We applied this technique to estimate the $PW$ relations for DCEP\_F,
DCEP\_1O and T2CEP, where this last sample includes only BLHER and
WVIR as did above for the LMC and SMC. 
In more detail, the fitting procedure has been carried out using the Nonlinear Least
  Square ({\it nls}) routine included in the {\it R}
  package\footnote{http://www.R-project.org/}. 
We adopted a weighted fitting conjugated with the bootstrap method to measure robust
errors on the parameters of the fit. In practice, the procedure is
repeated 1000 times (we increased the number of bootstraps until the
results were not depending on this number) and for each bootstrap we obtained a value of $\alpha$ and
$\beta$. The average values for these parameters and their standard deviations are obtained
from the resulting distributions. An example of the results is shown in
Fig.~\ref{figBootstrap}, where the distributions of $\alpha$ and
$\beta$ are reported, as well as that of the residuals around 
the ABL function. The results of the fitting procedure for the
different cases are shown in the first three rows of Tab.~\ref{resultsMW}.

%We adopted a weighed non
%linear fit based on Levenberg–Marquardt algorithm using the package
%{\it Gnuplot\footnote{http://www.gnuplot.info/}}. Our fitting
%procedure also foresees $\sigma$-clipping to exclude obvious outliers
%that can affect significantly the result. After several experiments,
%table results were found for a clipping level of 5 $\sigma$ for all
%the Cepheid types except for DCEP\_F for which we used 4 $\sigma$.
%The parameters of the $PW$ obtained with this procedure are shown in the
%lower part of Tab.~\ref{tabResults}, whereas the relationships are
%displayed with solid lines in the left panels of
%Fig.~\ref{figPLMW}. The goodness of the fits can be better appreciated
%in the bottom-left panel where we have only plotted objects with
%relative parallax error $<$20\%.  

A comparison of the $PW$ slopes between LMC and MW in
Tab.s~\ref{tabResults} and ~\ref{resultsMW} reveals that the slopes
of the $PW$ relations for the DCEP\_F  and T2CEPs are completely consistent one each
other within the errors, whereas for DCEP\_1O the discrepancy is 
of the order of 2 $\sigma$ level, being the slope of the MW sample
steeper than that of LMC. However the large error on the slope of the
MW sample makes this comparison not very stringent. 

Note that the low dependence on metallicity of the slope for
DCEP\_Fs is in agreement with previous works (both in theoretical and
observational) as it is generally found that the slope of the $PW$ for many different
band combinations has a very small dependence on the metallicity 
\citep[see e.g.][and references therein]{Fiorentino2007,Ngeow2012b,Dicriscienzo2013,Fiorentino2013,Gieren2018}.
We will come back on this argument in the next section. 

Similarly, for T2CEPs we do not find a significant dependence of the
slope of the $PW$ on the average metallicity of the parent
population, again in agreement with literature \citep[see
e.g.][]{Matsunaga2009,Matsunaga2011,Ripepi2015}. 

To the aim of comparing the zero points of the $PW$ relations
holding for MW and LMC, we imposed the proper values of $\beta$ for the LMC 
in Eq.~\ref{eqABL} and re-run the fitting procedure with the same
modality as before. The result of this operation is reported in the
second series of  three rows in Table~\ref{tabResults} and graphically in the right
panels of Fig.~\ref{figPLMW}. As expected the zero points of the
relations for DCEP\_Fs and T2CEPs are not significantly different than
in the previous case, whereas the contrary is true for DCEP\_1Os.
We will use these results in Sect.~\ref{LMC}.

As a final note, we underline that, due to the lack of thorough information in
the literature, in this work we are not considering
the source of uncertainty represented by the duplicity among DCEPs
whose incidence is highly uncertain, but estimated to be as large as
35-50\% or even more \citep[see][and references
therein]{Anderson2018}. The presence of companions for DCEPs affects
not only the parallaxes measured by {\it Gaia} (duplicity is not taken
into account in DR2), but also their photometry, thus possibly
representing a potential significant source of uncertainty. Next {\it
  Gaia} data releases will allow us to also face this important issue.

\subsection{PL relations in $G$, $G_{BP}$ and $G_{RP}$ bands for MW DCEPs}
\label{sectSelected}

To the aim of providing $PL$ relations in the {\it Gaia} $G$, $G_{BP}$
and $G_{RP}$ bands for the MW Cepheids, we need an estimate of the
reddening. As the {\it Gaia} DR2 does not include reliable interstellar
extinctions yet,  we have to use literature data. Thus, we have found
that reliable $E(B-V)$ values are available for a subsample of 301 objects classified as
DCEPs in Tab.~\ref{tabReclassified}. The main source for the reddening 
was \citet{Fernie1990}, whereas additional values were taken 
from \citet{Majaess2008,Ngeow2012a,Kashuba2016}.
Only a few objects possess reliable reddening estimates among MW T2CEPs, 
so that we did not try to calculate $PL$s for these objects.  
As for the metallicity, we used the results by \citet{Genovali2013,Genovali2014,Genovali2015}.  

The reddening values found in the literature are
listed in Tab.~\ref{tabReddening} together with the mode of pulsation
(268 and 33 DCEP\_Fs and DCEP\_1Os, respectively),  the metallicity
estimate, and the sources for reddening and metallicity, respectively
(last two columns).

Before proceeding, we have first to estimate the absorption in the
{\it Gaia} bands in terms of $E(B-V)$. To this aim we used again the
\citet{Jordi2010} tables, and, adopting the same procedure outlined in
Sect.~\ref{sectMCs}, we obtained starting values of ~2.90, 3.60 and
~2.15  for the ratios $A(G)/E(B-V)$, $A(G_{BP})/E(B-V)$ and
$A(G_{RP})/E(B-V)$, respectively. 

\begin{table*}
\caption{Reddening and metallicity for the 301 known MW Cepheids
  having {\it Gaia} DR2 intensity averaged magnitudes in the  $G$, $G_{BP}$ and
$G_{RP}$ bands coming from the  {\it Cepheids\&RRLyrae SOS} pipeline. The meaning of the
different columns is the following: (1) Literature name; (2) mode of
pulsation; (3) {\it Gaia} DR2 source identification; (4)-(5)  $E(B-V)$
and  error on its value; (6) metallicity ([Fe/H] value); (7)-(8) reference for
$E(B-V)$ and [Fe/H], respectively.
Note that the errors on metallicity
are not provided as usually not available object by object. They can
be estimated to be $\sim$0.1-0.15 dex.
The table is published in its entirety only in the electronic edition of the journal. 
A portion including the first 15 lines is shown here for guidance
regarding its form and content.} 
            % title of Table 
\label{tabReddening}      % is used to refer this table in the text 
\centering                          % used for centering table 
\begin{tabular}{l l c c c c c c}        % centered columns (4 columns) 
\hline\hline                 % inserts double horizontal lines 
\noalign{\smallskip}
 Name &  Mode  & Source\_id  & $E(B-V)$ &  $\sigma E(B-V)$ & [Fe/H]  &Ref1 & Ref2  \\
  &    &   & mag  &  mag & dex   & &   \\
(1) &  	(2) &  	(3) &  	(4) &  	(5) &  	(6) &  	(7) &  	(8) \\
\noalign{\smallskip} \hline \noalign{\smallskip}
     AA  Gem   &  DCEP\_F      &     3430067092837622272   &   0.380   &   0.019   &  -0.14   &    1 &  5   \\    
     AC  Cam   &  DCEP\_F      &      462252662762965120   &   0.915   &   0.046   &  -0.16   &    1 &  5   \\    
     AC  Mon   &  DCEP\_F      &     3050050207554658048   &   0.539   &   0.035   &  -0.06   &    1 &  5   \\    
     AD  Cam   &  DCEP\_F      &      462407693902385792   &   0.929   &   0.013   &  -0.28   &    1 &  5   \\    
     AD  Cru   &  DCEP\_F      &     6057514092119497472   &   0.681   &   0.013   &   0.08   &    1 &  5   \\    
     AD  Gem   &  DCEP\_F      &     3378049163365268608   &   0.173   &   0.019   &  -0.14   &    1 &  5   \\    
     AD  Pup   &  DCEP\_F      &     5614312705966204288   &   0.386   &   0.021   &  -0.06   &    1 &  5   \\    
     AE  Vel   &  DCEP\_F      &     5309174967720762496   &   0.735   &   0.058   &   0.11   &    1 &  5   \\    
     AG  Cru   &  DCEP\_F      &     6059635702888301952   &   0.257   &   0.021   &   0.05   &    1 &  5   \\    
     AH  Vel   &  DCEP\_1O     &     5519380077440172672   &   0.038   &   0.020   &   0.09   &    1 &  5   \\
\noalign{\smallskip} \hline \noalign{\smallskip}
\end{tabular}
\tablefoot{References: 1 = \citet{Fernie1990}; 2 = \citet{Ngeow2012a};
  3 = \citet{Majaess2008}; 4 = \citet{Kashuba2016}; 5 = \citet{Genovali2013,Genovali2014,Genovali2015}  }
\end{table*}     

\noindent  

We used again the ABL formulation of Eq.~\ref{eqABL} and the bootstrap technique to
  derive the relevant $PL/PW$ relations.  Now the observed magnitudes in
  the exponent of the right term can be the apparent Wesenheit $W$ or
  the observed de-reddened magnitudes $G^0$, $ G^0_{BP}$ and
  $G^0_{RP}$. With this formulation and the same
procedure of Sect.~\ref{sectPW} we calculated the $PL$ relations in the
{\it Gaia} $G$, $G_{BP}$ and $G_{RP}$ band for the MW DCEPs subsample
described above. Analysing the dispersion of the residuals, we checked
the above defined total-to-selective extinction ratios, by varying their values and re-estimating the
dispersion of the residuals (of the ABL) at any step. We retained the ratio values that returned the
smallest dispersions. They are shown in Eqs.~\ref{eqAbsorptions1} to 
~\ref{eqAbsorptions3}, where the uncertainties were estimated by looking
at the values of total-to-selective extinction ratios that produced an
increase in the dispersion. We remark that, owing to the
  large $G$, $G_{BP}$ and $G_{RP}$ band-widths, these total-to-selective
  ratios are only valid in the interval of colors spanned by Cepheids.

\begin{eqnarray}
A(G) & = & (2.70\pm0.05) E(B-V)  \label{eqAbsorptions1} \\
A(G_{BP}) & = & (3.50\pm0.10) E(B-V) \label{eqAbsorptions2} \\
A(G_{RP}) & = & (2.15\pm0.05) E(B-V) \label{eqAbsorptions3}
\label{eqAbsorptions}
\end{eqnarray}
\noindent
Finally, adopting the relations of Eqs.~\ref{eqAbsorptions1} to 
~\ref{eqAbsorptions3} we calculated the $PL$ relations in the {\it Gaia} $G$, $G_{BP}$ and $G_{RP}$ band
for the MW DCEPs subsample. 
The results are shown in the second part of Tab.~\ref{resultsMW} and
Fig.~\ref{figSelected}.  Note that we have also recalculated the $PW$ 
using the subsample adopted here. An inspection of
Tab.~\ref{resultsMW}  shows agreement within 1 $\sigma$ between
the $PW$s derived using the full sample and the subsample of
DCEP\_Fs. The same comparison is less meaningful for DCEP\_1Os because
of the huge errors, caused by the intrinsic large dispersion of the
full sample and by the small statistic in the case of the subsample. 

As for the subsample discussed in this section we have also available
the information about metallicity (see Tab.~\ref{tabReddening}), we
tried to derive $PW$ relations using the following ABL
definition including an additional term to take into account the dependence of the zero point on
the metallicity $[Fe/H]$ :

\begin{equation}
{\rm ABL}=10^{0.2 W_A}=10^{0.2(\alpha+\beta\log P +\gamma[Fe/H])}=\varpi10^{0.2W-2}
\label{eqABLZ}
\end{equation}

\noindent
where $W_A$ and $W$ are the absolute and relative Wesenheit
magnitudes, respectively.
In principle, also the $\beta$ term depends
on metallicity, but a comparison of the slopes for DCEP\_Fs in the
LMC (Tab.~\ref{tabResults}) and MW (first line of
Tab.~\ref{resultsMW}), shows that the dependence of $\beta$ on
metallicity can be expected reasonably low to be ignored. As this is
not true for the $PL$s, in the following we use only the
Wesenheit magnitudes.

Adopting the usual bootstrap technique applied to the ABL formulation
of Eq.~\ref{eqABLZ}, we obtain the result reported in the penultimate
line of Tab.~\ref{resultsMW} and Fig.~\ref{figFeh}. The derived 
 metallicity term $\gamma$=-0.237$\pm$0.199 dex/mag, even if only 
barely significant (~1$\sigma$), means that at fixed period and color, metal
poor stars are fainter. Note that these results are in good agreement
with \cite{Groenewegen2018} who derived $PL/PW$
relations in the optical and NIR bands adopting a subsample of
DCEPs with Gaia DR2 parallaxes and literature
photometry/spectroscopy and also with theoretical predictions for the
dependence of DCEP optical PW functions on metallicity
\citep[see Fig. 9 in][]{Caputo2000}.
Again, to compare the results for the MW and LMC, we recalculated the
ABL of Eq.~\ref{eqABLZ} but imposing the LMC value for the term
$\beta$. The outcome of this exercise is shown in the last line of
Tab.~\ref{resultsMW} and Fig~\ref{figFeh}. Not surprisingly, the
metallicity term becomes much less significant, as part of the metallicity dependence has
been absorbed by the variation of the slope. 

To obtain more stringent constraint on the dependence of DCEP  $PW$
and $PL$ relations 
on metallicity we will need not only more precise
parallaxes (expected in the next {\it Gaia} releases) but also to
increase the sample of DCEPs possessing accurate and homogeneous
measurements of metallicity by means of high resolution spectroscopy,
possibly extending the metallicity range spanned by the MW
DCEPs analysed here. 
In fact, only a few objects reach a metallicity value as low as that of the LMC
($[Fe/H]\sim$-0.4 dex), with the large majority of the pulsators
clustering around
$[Fe/H]\sim+0.05\pm0.1$ dex (see Tab.~\ref{tabReddening}).

\subsection{Distance of the LMC and zero points of the {\it Gaia} DR2
  parallaxes for Cepheids.}

\label{LMC}
 
In the previous sections we have estimated the $PW$ relations in the Gaia
bands for both the LMC and the MW using the slopes of the LMC. 
This operation makes it straightforward to estimate the distance of the
LMC, that is an important anchor for the extragalactic distance
scale, by comparing the zero points of the relative and absolute PWs
in the LMC and MW, respectively. We performed this exercise for
DCEP\_Fs and
T2CEPs as the $PW$ for DCEP\_1Os is too uncertain. For DCEP\_Fs we used
both the $PW$ without and with the metallicity term. In this last
case we adopted $[Fe/H]=-0.43$ dex for the LMC
\citep[]{Mucciarelli2011}, whereas for the MW we took the average of
the distribution of metallicities listed in Tab.~\ref{tabReddening},
i.e. $[Fe/H]=+0.05\pm0.13$ dex.
 The results are reported in the second column of  Tab.~\ref{tabLMC},
where the errors on the Distance Moduli (DMs) have been calculated
summing in quadrature the uncertainties on the zero points ($\alpha$
terms) and the metallicity ($\gamma$) when needed (see Tab.s~\ref{tabResults} and \ref{resultsMW}).
As a result, the DM$_{\rm LMC}$ obtained are always 
significantly  longer than the commonly accepted
value of $\sim$18.50 mag \citep[see
e.g.][]{Piet2013,Degrijs2014,Riess2018b}, even if the parallax zero point
correction of +0.046 mas by \citet[][]{Riess2018b} has already been applied. 

Conversely, if we use this value for the LMC distance as reference, we can recalculate the zero
point offset of the {\it Gaia} DR2 parallaxes, discovering that the
parallaxes zero point offset needed to obtain a DM$_{\rm LMC}\sim$18.50 is
of the order of +0.1 and +0.07 mas for the DCEP\_F  and T2CEPs ,
respectively. 
These results are in very good accordance with a similar analysis
carried out by \citet{Groenewegen2018} to which we remand the reader for a 
more detailed discussion.

\begin{table}
\caption{Results for the distance of LMC (see text).}             % title of Table
\label{tabLMC}      % is used to refer this table in the text
\centering                          % used for centering table
\begin{tabular}{l c cc }        % centered columns (4 columns)
\hline\hline                 % inserts double horizontal lines
\noalign{\smallskip} 
 Type & DM (mag)  &  $[Fe/H]$ term\\
\noalign{\smallskip} \hline \noalign{\smallskip}
 DCEP\_F   &   18.699$\pm$0.024  & no \\ 
DCEP\_F    &   18.673$\pm$0.085  &  yes \\ 
T2CEP        &   18.587$\pm$0.065  &   no \\
\noalign{\smallskip} \hline \noalign{\smallskip}
\end{tabular}
\end{table}

\section{Summary}

In this paper we have re-analysed the sample of Cepheids published in
the context of {\it Gaia} DR2 by \citet{Clementini2018}. The main
achievements of this paper are the following:

\begin{itemize}

\item
We have calculated the $PL/PW$ relations in the {\it Gaia} bands 
$G$, $G_{BP}$ and $G_{RP}$ for all the Cepheid types (DCEP, ACEP,
T2CEP) both in the LMC and SMC. These relations will be incorporated
in the next versions of the  {\it Gaia Cepheids\&RRLyrae SOS} pipeline adopted to classify the 
Cepheids in the {\it Gaia} DR3 \citep[see][]{Clementini2018}.

\item 
We carried out a careful re-analysis of the classification of the 2116
Cepheids of all types reported by \citet{Clementini2018} as belonging
to the MW. We first conducted a literature search for alternative
classification and period determination for these objects. Afterwards
we re-classified each object by visually inspecting its LC and position in
the $PW$ and Period-Fourier parameters. 

As a result, a total of 1257 stars have been classified as Cepheid of any type, 
84 objects as RR Lyrae and 647 as variables of other type (in addition
to the 128 stars with no classification).

Among these 1257 Cepheids, 713 were Cepheids already known in the
literature, 274 are new Cepheids completely unknown in the literature or indicated generically as
``variable'', and 270 objects were known in the literature with a
different classification. In total we classified 
800 DCEPs, 108 ACEPs and 336 T2CEPs, plus 13 Cepheids for which we were not able to find an
appropriate sub-classification in type. 

Among the MW sample we have individuated a total of 61 and 73 Cepheids 
of different types hosted by the LMC and SMC, 8 and 7 out of these
samples were not known in the literature as LMC/SMC objects. 

In addition, we were able to associate an ACEP\_F variable with  the URSA MINOR dwarf spheroidal galaxy, a WVIR pulsator 
with the GC NGC 6254 and a variable of unknown type to NGC 6266.

\item
Using the re-classified Cepheid sample, we used the ABL formulation to
derive $PW$ relations in the {\it Gaia} bands for the MW DCEP\_F, DCEP\_1O and T2CEP (BLHER and WVIR).  
The use of the ABL formulation allows us to derive slopes and zero
points for the $PW$ that are almost unbiased, as we did not do any
kind of selection on the sample. 
The adoption of a subsample (301 objects) of well characterised MW
DCEPs 
possessing reliable reddening and metallicity estimates, allowed us to
calculate also the $PL$ relations for the  $G$, $G_{BP}$ and $G_{RP}$
bands for DCEP\_F and DCEP\_1O. 

In addition, using the quoted subsample, we were able to 
  investigate for the first time the dependence on metallicity of the
  $PW$ relation for DCEP\_Fs in the {\it Gaia} bands. 
As a result, we have derived a modestly significant
(~1$\sigma$) dependence ($\gamma$=-0.237$\pm$0.199 dex/mag), in the
sense that at fixed period metal poor stars are fainter. More precise
parallaxes and spectroscopic measures will be needed to address firmly
this point.

\item
We calculated also the $PW$  relations for the MW by imposing the slope of the PW
relations in the LMC and redetermining the zeropoints. 
By comparing the relative zeropoints between the
MW and the LMC $PW$ for DCEP\_F and T2CEP, we obtained two different
estimates of the LMC distance. These values are larger
than the usually accepted value for the LMC DM$\sim$18.50. To
reconcile the results found here with the latter we need to increase
the zero points of the {\it Gaia} DR2 parallax by at least 0.07 mas,
in agreement with recent literature results.

\end{itemize}

The {\it Gaia} DR2 photometry and parallax for Cepheids in the MW allowed a
significant step forward in the classification of the different type
of Cepheids. Indeed, the excellent photometric quality, even
conjugated the relatively low-accurate parallaxes for the
sample of objects discussed in this paper, allowed us to revise the
literature classification for more than two hundred objects.     

In conclusion, without entering in details beyond the scope of present
paper, the results presented in this work seem to confirm the
\citet{Groenewegen2018} suggestion that the parallaxes for MW Cepheids 
in the {\it Gaia} DR2, appear still too uncertain to allow a
significative decrease of the error on the value of $\rm H_0$.
To this we have to add the uncertainties on the extinction law, on the
impact of metallicity, as well as on binarity, affecting
both astrometry and photometry of Cepheids.

Great improvements are awaited from the next {\it Gaia} DR3 and DR4
for all these issues. Indeed, these releases are expected to present 
extremely accurate photometry and astrometry corrected for
the effect of multiplicity, as well as individual information on
reddening, metallicity and duplicity for a large fraction of the sky.   
Therefore, DR3 and DR4 will certainly allow to make consistent steps forward in the
accuracy of the extragalactic distance scale, helping to reduce the uncertainty on the value
of $\rm H_0$ to less than 1\%.

\begin{acknowledgements}
We thanks our anonymous referee whose comments helped to improve the paper.
We gratefully thank G. Clementini for stimulating discussions  on the
subject of this paper.
This work has made use of data from the European Space Agency (ESA) mission
{\it Gaia} (\url{https://www.cosmos.esa.int/gaia}), processed by the {\it Gaia}
Data Processing and Analysis Consortium (DPAC,
\url{https://www.cosmos.esa.int/web/gaia/dpac/consortium}). Funding for the DPAC
has been provided by national institutions, in particular the institutions
participating in the {\it Gaia} Multilateral Agreement.
In particular, the Italian participation
in DPAC has been supported by Istituto Nazionale di Astrofisica
(INAF) and the Agenzia Spaziale Italiana (ASI) through grants I/037/08/0,
I/058/10/0, 2014-025-R.0, and 2014-025-R.1.2015 to INAF (PI M.G. Lattanzi).

This research has made use of the International Variable Star Index
(VSX) database, operated at AAVSO, Cambridge, Massachusetts, USA.
This research has made use of the SIMBAD database,
operated at CDS, Strasbourg, France 

\end{acknowledgements}


\begin{appendix} %First appendix

\section{Acronyms for the literature variability types}

In Tab.~\ref{tabAcronyms} we expand the variability types adopted for Tab.~\ref{tabLiterature}.

\begin{table}
\caption{Acronyms adopted in Tab.~\ref{tabLiterature} to indicate the
different variability types.}\label{tabAcronyms}
\centering
\begin{tabular}{ll} 
\hline\hline             
\noalign{\smallskip}
 Acronym & Definition \\
\hline
ACEP\_F & Anomalous Cepheids Fundamental mode \\  
ACEP\_1O & Anomalous Cepheids First Overtone \\  
AGB    &     AGB Star      \\  
AGN    &     Active Galactic Nuclei      \\  
BLHER    &     Type II Cepheid BL Herculis      \\ 
BLLac    &     BL Lacertae-type object      \\ 
Be    &      Be eruptive stars      \\ 
Carbon    &     Carbon star \\
CV    &     Cataclysmic variable      \\ 
DCEP    &     Delta Cepheid      \\ 
DCEP\_1O    &     Delta Cepheid First Overtone      \\ 
DCEP\_2O    &     Delta Cepheid Second Overtone      \\ 
DCEP\_F    &     Delta Cepheid Fundamental mode      \\ 
EB    &     Beta Lyrae-type eclipsing systems.      \\ 
EB    &     Eclipsing Binary      \\ 
EC    &     Contact binaries      \\ 
ELL    &     Rotating ellipsoidal variables      \\ 
Em    &     Emission Line star      \\ 
Eruptive    &     Eruptive      \\   
ErupIRR    &     Eruptive Irregular      \\  
FUOri     &     Fu Orionis Type star      \\ 
HB    &     Horizontal Branch star      \\ 
HS    &     Hot Subdwarf star      \\ 
Irr    &     Irregular      \\ 
LPV    &     Long Period Variable      \\ 
Mira    &     Variable Star of Mira Cet type      \\ 
NC    &     Not classified      \\ 
Orion    &     Variable Star of Orion Type      \\ 
PostAGB    &     Post AGB star      \\ 
Puls    &     Pulsating Variable star.      \\ 
RC    &     Rapid Change      \\ 
RG   &     Red Giant      \\ 
ROT    &     Rotational      \\ 
RR    &     RR Lyrae      \\ 
RRab    &     RR Lyrae type ab      \\ 
RRc    &     RR Lyrae type c      \\ 
RSCVn &  RS Canum Venaticorum type      \\ 
RVTAU    &      RV Tauri type      \\ 
SARG    &     Small Amplitude Red Giant      \\ 
SARG\_A    &     Small Amplitude Red Giant, subclass A      \\ 
SARG\_B    &     Small Amplitude Red Giant, subclass AB      \\ 
Semireg    &     Semiregular      \\ 
SXPHE    &     Sx Phoenicis star      \\ 
T2CEP    &     Type II Cepheid      \\ 
TTAU/CTTS    &     T Tauri star/Classical T Tauri Stars      \\ 
UXOri    &     UX Orionis Type star      \\ 
VAR    &     Variable      \\ 
WR    &     Wolf Rayet      \\ 
WUma    &     W Uma      \\ 
WVIR    &     Type II Cepheid W Virginis      \\ 
XRB    &     X Ray Binary \\
YSO    &    Young Stellar Object \\
\noalign{\smallskip} \hline \noalign{\smallskip}
\end{tabular}
\end{table}

\end{appendix}

\end{document}